\documentclass[onecolumn]{emulateapj}

\usepackage{natbib}
\usepackage{graphicx,amssymb}

\usepackage[dvipsnames]{color}

\shorttitle{Supernovae and AGN driven galactic outflows}
\shortauthors{Sharma and Nath}

\begin{document}

\newcommand{\half}{\frac{1}{2}}
\newcommand{\3}{\ss}
\newcommand{\n}{\noindent}
\newcommand{\eps}{\varepsilon}
\def\be{\begin{equation}}
\def\ee{\end{equation}}
\def\ba{\begin{eqnarray}}
\def\ea{\end{eqnarray}}
\def\de{\partial}
\def\msun{M_\odot}
\def\div{\nabla\cdot}
\def\grad{\nabla}
\def\rot{\nabla\times}
\def\ltsima{$\; \buildrel < \over \sim \;$}
\def\simlt{\lower.5ex\hbox{\ltsima}}
\def\gtsima{$\; \buildrel > \over \sim \;$}
\def\simgt{\lower.5ex\hbox{\gtsima}}
\def\etal{{et al.\ }}
\def\red{\textcolor{red}} 
\def\blue{\textcolor{blue}}
\def\del{\partial}
\newcommand{\pd}{\partial}

\newcommand\redc[1]{{\color{red} \bf #1}}

\newcommand{\dvr}{\frac{dv}{dr}}
\newcommand{\dcsr}{\frac{dc_s^2}{dr}}
\newcommand{\dphr}{\frac{d\Phi(r)}{dr}}
\newcommand{\dMr}{\frac{d\mathcal{M}^2}{dr}}

\newcommand{\ddr}{\frac{d}{dr}}
\newcommand{\ddphi}{\frac{d}{d\phi}}
\newcommand{\DDR}{\frac{\partial}{\partial{r}}}
\newcommand{\DDphi}{\frac{\partial}{\partial{\phi}}}
\newcommand{\DDZ}{\frac{\partial}{\partial{z}}}
\newcommand{\DDt}{\frac{\partial}{\partial{t}}}
\newcommand{\der}[2]{\frac{d{#1}}{d{#2}}}
\newcommand{\DER}[2]{\frac{\partial{#1}}{\partial{#2}}}
\newcommand{\pdt}{\DER{}{t}}
\newcommand{\bO}{\boldsymbol \omega}
\newcommand{\bnab}{\boldsymbol \nabla}
\newcommand{\ro}{r_{0}}
\newcommand{\rp}{r\prime}
\newcommand{\pp}{\phi\prime}
\newcommand{\mach}{\mathcal{M}}
\newcommand{\Msun}{{\rm M}_\odot}
\newcommand{\mbh}{M_{\bullet}}
\newcommand{\vbh}{v_{\bullet}}
\newcommand{\mhalo}{M_{200}}
\newcommand{\albe}{{\alpha/\beta}}
\newcommand{\sfr}{\dot M_{\star}}
\newcommand{\superdel}{\mathcal{M}^{{2\over\gamma-1}}\left(\frac{\gamma-1+2/\mathcal{M}^2}{\gamma+1}\right)^{\frac{\gamma+1}{2(\gamma-1)}}}
\newcommand{\subdel}{\left(\frac{3\gamma+1/\mach^2}{3\gamma+1}\right)^{-\frac{3\gamma+1}{5\gamma+1}} \left(\frac{\gamma-1+2/\mach^2}{\gamma+1}\right)^{\frac{\gamma+1}{2(5\gamma+1)}}}
\newcommand{\kms}{km s$^{-1}$}
\newcommand{\Mvir}{M_{\rm vir}}
\newcommand{\rvir}{r_{\rm vir}}
\newcommand{\zvir}{ z_{\rm vir}}
\newcommand{\vvir}{v_{c}}

\title{SUPERNOVAE AND AGN DRIVEN GALACTIC OUTFLOWS
}

\author{Mahavir Sharma and Biman B. Nath}
\affil{Raman Research Institute, Sadashiva Nagar, Bangalore 560080, India}
\email{mahavir@rri.res.in; biman@rri.res.in}

\begin{abstract}
We present analytical solutions for winds from galaxies with NFW dark matter halo. We consider winds driven by energy and mass injection from multiple supernovae (SNe), as well as momentum injection due to radiation from a central black hole. We find that the wind dynamics depends on three velocity scales: (a) $v_\star \sim (\dot{E} / 2 \dot{M})^{1/2}$ describes the effect of starburst activity, with $\dot{E}, \dot{M}$ as energy and mass injection rate in a central region of radius $R$; (b) $\vbh \sim (G\mbh / 2 R)^{1/2}$ for the effect of a central black hole of mass $\mbh$ on gas at distance $R$ and (c) $v_{s} =(GM_h  / 2\mathcal{C}r_s)^{1/2}$ which is closely related to the circular speed ($\vvir$) for NFW halo, with $r_s$ as the halo scale radius and $\mathcal{C}$ is a function of halo concentration parameter. Our generalized formalism, in which we treat both energy and momentum injection from starbursts and radiation from central active galactic nucleus (AGN), allows us to estimate the wind terminal speed to 
be $(4v_\star^2 +6(\Gamma -1) \vbh^2 -4v_s^2)^{1/2}$, where $\Gamma$ is the
ratio of force due to radiation pressure to gravity of the central black hole.
Our dynamical model also predicts the following: (a) winds from quiescent star forming galaxies 
cannot escape from $10^{11.5} \le M_h \le 10^{12.5}$ M$_\sun$ galaxies, (b) circumgalactic gas at large distances
from galaxies should be present for galaxies in this mass range, 
(c) for an escaping wind, the wind speed in low to intermediate mass 
galaxies is 
$\sim 400\hbox{--} 1000$  km/s, consistent with observed X-ray temperatures;
 (d) winds from massive galaxies with AGN at Eddington limit have  speeds $\gtrsim 1000$ km/s.
We also find that the ratio $[2 v_\star ^2 -(1 -\Gamma)
\vbh^2] / \vvir^2$ dictates the amount of gas lost through winds.  
Used in conjunction with an appropriate relation between $\mbh$ and $M_h$, and an appropriate opacity of dust grains in infrared (K band), this ratio has the attractive property
of being minimum at a certain halo mass scale ($M_h \sim 10^{12\hbox{--}12.5}$ M$_\sun$) that signifies the cross-over of AGN domination in outflow properties from starburst activity at lower masses.
We find that stellar mass for massive galaxies scales as $M_\star \propto M_h^{0.26}$,
and for low mass galaxies, $M_\star \propto M_h^{5/3}$.
\end{abstract}


\section{Introduction}
In the standard scenario of galaxy formation the baryonic matter falls inside the potential  wells created by  dark matter halos. This in-falling material cools and forms stars. This picture is met with problems as the galaxy stellar mass function does not follow the halo mass function. Both  low and  high mass halos have significantly lower than predicted value of stellar masses \citep{Somerville06,Moster10,Behroozi10}.
To reconcile with these problems it has been proposed  that the  SNe and starbursts provide thermal energy injection and cause large amount of mass to flow out of the galaxy as galactic superwinds \citep{Dekel86,larson74,Oppenheimer06}. The star formation is suppressed as the galaxies lose a significant portion of their baryons due to this negative feedback. Although this picture can provide explanation for the low ratios of $M_\star/M_h$ for the low mass halos, but for the high mass halos, gravity becomes  strong and the SNe  are not sufficient to drive the gas out. In order to resolve the discrepancy at high mass end, it has  been argued that the AGN outflows may sweep away the baryons and suppress the star formation in high mass galaxies \citep{Silk98,Wyithe03,Matteo05,Springel05,Croton06,Bower06}. These two feedback processes, when considered together, are believed to explain the shape of galaxy stellar mass function at both low and high mass end \citep{Binney04, Cattaneo06, Puchwein12}.

Apart from their cosmological importance as a feedback process, galactic winds have been a topic of research as a gas dynamical problem in galactic physics. Speculations on the possibility that galaxies can harbour large scale winds followed the models of solar wind 
developed by \cite{Parker65}. \cite{Burke1968} proposed a model of trans-sonic winds from the galaxy 
with heat and mass addition from SNe. It was further proposed that galactic winds may cause  ellipticals to lose all of their gas \citep{Johnson71,Mathews71}. The review by \cite{Holzer70} gives an elaborate account of the theoretical aspects of solar and galactic winds. 
\citet{CC85} showed that energy injection at the center can drive  a fast super-wind from the dwarf star-burst galaxy M82. In this work the gravity of the galaxy was not considered and the obtained solutions were trans-sonic with a heat injection up to the sonic point. \cite{Wang95} modeled the wind from a power law gravitational potential and showed that wind may escape the galaxy or settle in a galactic corona depending on the mass of the galaxy and the effect of radiative cooling. \cite{Silich11} studied the effects of cooling  on winds from individual star clusters with exponential stellar density distribution. Winds driven by cosmic rays have also been studied in the literature \citep{Ipavich1975,Breitschwerdt1991,Samui2008,Uhlig12}.

Observations show that the winds do not consist of a homogeneous medium. The hot gas emitting X-rays and the cold/warm gas visible in  emmision and absorption lines \citep{Strickland04,Bouche12,Kornei12}, coexist in the galactic winds \citep{Heckman00,Veilleux05}. Hot phase appears as a smooth flow of tenuous gas while the colder phase is clumpy in nature. It is usually believed that
 cold neutral clouds form as a result of thermal instabilities in the hot flow and  they are entrained with the parent flow because of its ram pressure \citep{Heckman93}. Hydrodynamical simulations with radiative cooling also supported this scenario \citep{Suchkov94,Strickland00,Cooper08}.  However, recent observations show that the velocity of neutral clouds does not correlate with the velocity of  hot flow and rather it correlates with the circular speed of the host galaxy \citep{Martin2005,Rupke05}. If one considers the momentum driven winds where the radiation from the galaxy acts on the dust grains,  then these observations can be explained \citep{MQT05,Sharmaetal12,Chatto12,NathSilk2009}. However the radiation may not be sufficient in fainter low mass galaxies and ram pressure is still required to explain the cold winds in these galaxies \citep{SharmaNath12,Murray11,Hopkins12}. On the other hand in high mass ULIRGs,  radiation from AGN may be an alternate mechanism for driving outflows apart from SNe and stellar radiation.

In spite of the general consensus about the AGN driving in quasar outflows, observationally it has been hard to establish that outflows in galaxies are also powered by AGN \citep{RupkeAGN05,West12}.   However, recent observations do show compelling evidence for the AGN driving in galactic outflows \citep{Sturm11,Rupke11,Alexander10,Morganti07,Dunn10,Feruglio10,Fu09,Villar11}.  
From the theoretical point of view, considerable amount of work has been carried out in modeling radiation driven outflows in the immediate vicinity  of the AGN \citep{Murray95,Proga00,Kurosawa09,Risaliti10}. However, the effects of radiation from accreting black holes has not been discussed for driving galaxy scale outflows. \cite{MQT05} proposed the existence of a critical luminosity for the AGN (or the galaxy) necessary for the blow-out of all the available gas. The value of this critical luminosity depends on the  dust scattering opacity at UV. \citet{Everett07} studied Parker wind from AGN occurring at scales of $\sim100$ pc. \cite{Debuhr12} carried out simulations showing that the initial momentum injection and the fast outflow in the vicinity of hole may shock the surrounding ISM  and can result in a galaxy scale outflow.  Whether the outflows due to the AGN are energy conserving or momentum conserving has been a topic of debate as well. \cite{Silk10} argued that the energy driven outflows are not possible in the galactic bulges. \cite{King11} proposed that AGN outflows are momentum driven at small scales and energy driven on larger scales. \cite{Faucher12} showed that the cooling in the region, shocked by AGN radiation pressure, may not be effective and the outflows can be energy conserving. \cite{McQ12} studied the large scale motion of momentum-conserving supershells from a dark matter halo. \cite{Novak12} carried out a radiation transfer calculation assessing the efficiency of various components of AGN spectrum in driving outflows. This work showed that  most of the UV flux is quickly absorbed and re-radiated in IR.  The IR radiation can drive a dusty outflow and may result in mass loss much higher than the line driving mechanisms,  on the scales connecting the AGN and host galaxy as shown by radiation hydrodynamic simulations \citep{Dorod12} .

It is evident from the  studies on starburst and AGN  driven outflows that these processes play an important role in the formation and evolution of galaxies. In  galaxy formation models, these two processes are generally invoked using simple recipes through feedback factors. However, in theoretical models of winds they have been treated separately for the low mass and high mass galaxies.  There is a lack of models that develop a complete hydrodynamic theory of winds which can envisage both these feedback processes.  In the present paper we address this problem analytically and bring in both these wind driving agents together and using the analytical results from our calculation we also attempt an explanation for the galaxy stellar to halo mass relation.

We start with the derivation of a general wind equation in \S 2, which accounts for any possible mass, energy and momentum injection. We then present a brief derivation of  SNe driven wind model of CC85 and extend this model to the case including a dark matter halo in \S 3. Afterwards in \S 5 we introduce the momentum injection from the AGN and derive a general analytic solution for the galactic wind, which has inputs from SNe injection, NFW gravity and the central black hole. This solution leads us to important results like the terminal velocity of winds, condition for escaping winds and the dependence of wind properties  on the halo mass and the black hole mass. In \S 5.2 we show the velocity, density and temperature of the outflow as a function of the distance from the center. Interestingly a class of our solutions can explain the gas reservoirs in the the galactic halos seen in  observations  \citep{Tumlinson12} and simulations \citep{Vandevoort12,Stinson12}. 
 In \S 5.3 we study the cosmological implications of our results. We derive the scaling relation between  stellar ($M_\star$) and halo mass ($M_h$) using simple recipes and inputs from our models. We compare our analytically derived stellar to halo mass ratio (SHMR) with the observations and results of abundance matching. We discuss our results in \S 6.

\section{Basic  equations}
We consider steady spherically symmetric winds driven by mass and energy/momentum injection from processes that are confined in a central region of radius $R$. Consider a heating rate $\dot E$ and mass injection rate $\dot M$
in this region, which has a size of a few hundred parsecs (see below) then the basic fluid equations can be written as, 
\begin{eqnarray}
&& {1 \over r^2}\frac{d}{dr}(\rho v r^2) = \dot m = {\dot M \over V}\\
&& v \frac{dv}{dr} = -{1\over\rho}{dp \over dr}-{d\Phi\over dr} + f(r)-{\dot m v \over \rho}\\
&& {1\over r^2}{d\over dr}\left[\rho v r^2\left(\frac{v^2}{2}+{c_s^2\over \gamma - 1}\right)\right]+\rho v\left(\dphr-f(r)\right) = q =  {\dot{E}\over V} 
\end{eqnarray}
Here $V$ is the volume of the central region in which the energy injection and the mass injection is occurring. $\Phi(r)$ represents the gravitational potential and $f(r)$ is the momentum injection force per unit mass. $c_s$ is the Laplacian sound speed. In this work we will be considering the momentum injection from the AGN in optically thin limit hence the $f(r)$ has an inverse square dependence on $r$.  The above written system of equations do not have a critical point if the heating and mass 
injection is zero. However, for a finite energy and mass injection there is a critical point. For an extended energy and mass distribution the critical point can be determined numerically as done for super star-cluster winds in \citet{Silich11}. To extract maximum information analytically  we have considered the energy and mass injection to be confined in a region of radius $r=R$ following \citet[hereafter CC85]{CC85}.  Therefore in the present case the critical point is situated right at the boundary of central injection region.
 
In this work we do not consider the radiative cooling,
as it is generally believed 
that the energy loss via radiation  over the entire wind is small and less dominant than the adiabatic loss \citep{Grimes09}, hence the cooling does not affect the dynamics of the flow. However it may still be important for the thermodynamics of the flow (see appendix D) and may result in the precipitation of the wind which can not be dealt with the steady stable flow solutions. The dynamics and survival of  clouds formed by thermal instability in the galactic wind is also an important issue and it has been studied elsewhere in the category of cold winds \citep{Murray11,SharmaNath12,Marcolini05,Cooper09}. Therefore, we distinguish our model from that of the clumpy winds and  in the present work we  study analytically the large scale dynamics of homogeneous steady outflow from a NFW dark matter halo and its cosmological implications.

By introducing the mach number $\mach=v/c_s$, the above  equations can be transformed to the following ordinary differential equation (see Appendix A for a complete derivation).
\be
\frac{\mach^2-1}{\mach^2(\mach^2(\gamma-1)+2)}\dMr = {2\over r}-(1+\gamma\mach^2){\dot m \over \rho v}-\frac{\dot m(1+\gamma\mach^2)}{2 \rho v}\left(\frac{\dot E/\dot M}{\epsilon(r)}-1\right) + \frac{(\gamma+1)\left(f(r)-\dphr\right)}{2(\gamma-1)\epsilon(r)}
\label{central}
\ee
where $\dot{m}=\dot{M}/V$ and $\epsilon(r) = \frac{v^2}{2} + \frac{c_s^2}{\gamma-1}$. We will use $\gamma=5/3$. 
As mentioned above, the terms $\dot E$ and $\dot M$  represent the energy and mass injection in a central region of size $r=R$, beyond which they become zero. We  use $R=200$ pc for the present work.

\subsection{Zero gravity case: Chevalier \& Clegg's solution}
In this section we briefly reproduce the CC85 solution, therefore this subsection also serves as a consistency check for equation \ref{central}. In the CC85 solution gravitational force of the galaxy is not considered as the wind speeds were of the order of thousand \kms which is an order of magnitude  larger than the circular speed of starburst galaxy M82. There is no external driving force ($f(r)=0$). Hence the main driving force is the energy injection from SNe in a central $200$ pc region. The size of the central region also marks the boundary where the energy injection and the subsonic part ends.
 The wind equation then can be solved  analytically  for subsonic ($\mach<1$) and supersonic ($\mach>1$) part of the wind.

In order to derive the subsonic part of the solution, we can use $\dot m  = {3\rho v / r}$ which results from the integration of  continuity equation. 
Therefore equation \ref{central} becomes
 \be
\frac{\mach^2-1}{\mach^2(\mach^2(\gamma-1)+2)}\dMr = {2\over r}-{3(1+\gamma\mach^2) \over r}-\frac{3(1+\gamma\mach^2)}{2\ r}\left(\frac{\dot E/\dot M}{\epsilon(r)}-1\right) 
\label{CCpart1}
\ee
Direct integration of energy equation 3 by retaining the energy and mass injection  and neglecting the gravity and external driving results in, 
 $
 \epsilon(r) = {\dot E / \dot M} =2 v_\star ^2 \,,
 $
where we have defined a velocity parameter, $v_\star=\sqrt{\dot E/2\dot M}$ . It can be shown that at the critical point, $v_{\rm crit} = v(R) = c_s(R) = v_\star$. 
Substituting this $\epsilon(r)$ in equation  \ref{CCpart1}, we get,
\begin{eqnarray}
 \frac{\mach^2-1}{\mach^2 [(\gamma -1)\mach^2+2]}\frac{d\mach^2}{dr} =
 \frac{-1-3\gamma\mach^2}{r}
\end{eqnarray}
This can be integrated to get the following solution,
\be
\delta_<(\mach) =
 \subdel= \frac{r}{R}\ ; \quad  \quad  r<R
\label{defdel1}
\ee

Following similar steps and additionally dropping the injection terms, we can  evaluate supersonic part of the solution. Analytically it implies setting $\Phi=f =  \dot M = \dot E = 0$ in the wind equation \ref{central}. Therefore, we are left with the following differential equation,
 \begin{eqnarray}
 \frac{\mach^2-1}{\mach^2 [(\gamma -1)\mach^2+2]}\frac{d\mach^2}{dr} =
 \frac{2}{r}
 \label{CCpart2}
\end{eqnarray}
 This can be integrated to get the following solution for the supersonic part of the wind,
 \be
\delta_>(\mach) = \superdel=\left(\frac{r}{R}\right)^2 \ ; \quad  \quad r>R
\label{defdel2}
\ee
These expressions were arrived at by CC85. We shall use these definitions of  $\delta_<(\mach)$ and $\delta_>(\mach)$ in the rest of the paper. The terminal velocity of the wind in this solution can be obtained easily from the energy equation which gives $\epsilon(r) = v^2/2 + c_s^2/(\gamma-1)=2v_\star^2$. For $r\rightarrow\infty$, the sound speed can be neglected and we get $v_\infty=2v_\star = (2\dot E/\dot M)^{1/2}$ . In the supersonic part of the wind for $\mach>1$  the relation in equation \ref{defdel2} can be approximated by $\mach^3\propto r^2$. Therefore,  when the velocity attains its terminal value, then $c_s^2\propto T \propto  r^{-4/3}$ and $\rho\propto r^{-2}$.

\section{SN driven winds from NFW halo} 
In this section we  study the effects of NFW dark matter halo on the wind velocity in detail. One can therefore consider the calculation in this section as an extension of the supersonic part of CC85 solution. We proceed  first by fixing the dark matter halo parameters in the next subsection.
\subsection{Dark matter halo properties} \label{DMhalo}
Here we discuss the properties of the NFW dark matter halo which will be used throughout the rest of the paper. We consider a Navarro Frank and White(NFW) dark matter halo with a density profile $\rho = \rho_sr_s^3/r(r+r_s)^2$ \citep{NFW96}. For a  dark matter halo of total mass $M_h$,  which collapses at a redshift $z$,  the virial radius is given by
$\rvir = [3M_h/( 4 \pi \Delta_c(z) \Omega_m^z \rho_{crit})]^{1/3} (1+z)^{-1}$
Where $\Delta_c$ is the critical over density which can be written as  $\Delta_c(z) = 18\pi^2 -39 d_z^2 +82 d_z$, with $d_z = \Omega_m^z-1$ and $\Omega_m^z = \frac{\Omega_m(1+z)^3}{\Omega_m(1+z)^3+\Omega_\Lambda+\Omega_k(1+z)^2}$ with parameters having their usual meaning \citep{Bryan98,Bullock01}. 
We set  $\Omega_m = 0.258,\ \Omega_\Lambda = 0.742,\ h=0.72$, according to the results from WMAP5 \citep{Komatsu09}. Using these values, the virial radius can be written as 
\be
\rvir = 210 \ \left({M_h\over 10^{12}\ \Msun}\right)^{1/3}\left[{\Omega_m\over\Omega_m^z}{\Delta_c(z)\over 18\pi^2}\right]^{-1/3}(1+z)^{-1}\ \ {\rm kpc}
\label{defrvir}
\ee 
The corresponding  circular speed is, $\vvir = \sqrt{G M_h/\rvir}$. So it can be written as,
\be
\vvir = 143 \ \left({M_h\over 10^{12}\ \Msun}\right)^{1/3}\left[{\Omega_m\over\Omega_m^z}{\Delta_c(z)\over 18\pi^2}\right]^{1/6}(1+z)^{1/2}\ \ {\rm km\ s^{-1}}
\label{defvc}
\ee 
The gravitational potential due to the NFW dark matter halo can be written as,
\be
\Phi_{\rm NFW} = -{G M_h \over \ln(1+c)-c/(1+c)}\  {\ln(1+r/r_s)\over r} = -2v_s^2 \ {\ln(1+r/r_s)\over r/r_s}
\label{potNFW}  
\ee
In this $v_s^2 = GM_h/2\mathcal{C}r_s$ and  $\mathcal{C} = \ln(1+c)-c/(1+c) $. Also  $r_s = \rvir/c$ is the NFW scale radius with $c$ as the halo concentration parameter which depends  on both halo mass and the redshift of virialization. We use the fitting formula for $c$ given by \cite{Munoz11}.

\subsection{Effect of dark matter halo on winds}
We study the effect of gravity in the supersonic part of the solution. In the subsonic part, the NFW gravity does not make a difference as shown in Appendix B. Therefore the subsonic part is same as in the previous section and we can assume that  the energy injection propels the gas to a speed of $v_{\rm crit} = v_\star = (\dot E/2\dot M)^{1/2}$ at the critical point.  
To study the supersonic solution beyond the critical point ($r>R$) , we set $\dot E$ and $\dot m$ equal to zero, and use the NFW potential in equation \ref{central}, which results in the following equation for the supersonic part,
\begin{eqnarray}
 \frac{\mach^2-1}{\mach^2 [(\gamma -1)\mach^2+2]}\frac{d\mach^2}{dr} =
 \frac{2}{r} - \frac{2}{\epsilon(r)}\frac{d\Phi_{\rm NFW}(r)}{dr} 
 \label{CC2NFW}
\end{eqnarray}
where $\epsilon(r)={v^2/ 2} +{c_s^2 / (\gamma-1)}$.
From the direct integration of energy equation we obtain
\begin{eqnarray}
  &&\left. {v^2 \over 2}+{c_s^2 \over \gamma-1} + \Phi_{\rm NFW}(r) \right|_{R}^{r}= 0
  \nonumber \\
 \Rightarrow && \epsilon(r) = 2v_\star^2 + \Phi_{\rm NFW}(R) - \Phi_{\rm NFW}(r)
\end{eqnarray}
where we have used the value of velocity at the critical point as $v_{\rm crit} = v(R) = c_s(R) = v_\star = (\dot E/2\dot M)^{1/2}$ Substitution of  this $\epsilon(r)$ in equation \ref{CC2NFW} followed by integration results in,
\be
\ln|\delta_>(\mach)| = 2 \ln |r| + 2 \ln |2v_\star^2 + \Phi_{\rm NFW}(R) - \Phi_{\rm NFW}(r)| + const.
\ee
where $\delta_>(\mach)$ is defined in equation \ref{defdel2}. In Appendix C we have plotted the contours of $\mach$ versus r for an arbitrary value of constant.  For the analysis here we pick the solution which has a critical point at radius $R$.  
To pick that solution we set, $\delta_>(\mach)=1$ and $r=R$ in the above equation which fixes the constant and we get,
\be
\delta_>(\mach) = \left[{r\over R}\right]^2 \left[\frac{2v_\star^2 + \Phi_{\rm NFW}(R) - \Phi_{\rm NFW}(r)}{2v_\star^2}\right]^2
\ee

Using the definition of $\Phi_{\rm NFW}$, in equation \ref{potNFW} we get,
\be
\delta_>(\mach) = \frac{(r/R)^2}{2v_\star^2}  \left[{2v_\star^2 - 2v_s^2\left(\frac{\ln(1+R/r_s)}{R/r_s}-\frac{\ln(1+r/r_s)}{r/r_s}\right)}\right]^2
\simeq (r/R)^2 \left[ 1- {v_s^2 \over v_\star^2}\left(1-{\ln(1+r/r_s)\over r/r_s}\right) \right]^2
\ee
where  $v_s = \sqrt{GM_h/2\mathcal{C}r_s}$.
Comparing with eqn \ref{defdel2}, one can clearly see that the solution picks up an additional term due to the gravitational force of NFW halo.
The solution has three different regimes depending on the value of the ratio $v_\star^2/v_s^2$, as described below\\
\begin{itemize}
\item case 1 : $v_\star^2 >> v_s^2$ \\
This is the case when the initial injection upto the critical point is very strong and hence the wind velocity  
at the critical point is  large. Then we can neglect the ratio $v_s^2/v_\star^2$, and we get 
\be
\delta_>(\mach) = (r/R)^2
\ee
The NFW halo has negligible effect in this particular case  and we have recovered the supersonic part of CC85 solution.

\item case 2 : $v_\star^2  = v_s^2$ \\
In this case we get 
\be
\delta_>(\mach) = \left(\frac{r_s}{R}\  \ln (1+r/r_s)\right)^2
\ee
which shows that the gravity of dark matter halo affects the flow significantly.  As there is a logarithmic dependence in the above equation, therefore in large $\mach$ limit, the Mach number and hence the velocity increases slowly with r, in comparison to the case without gravity. 

\item case 3 : $v_\star^2 < v_s^2$ \\
In this interesting case, we find that the wind speed decreases and finally becomes zero at some distance.
This distance is decided by the ratio   $v_\star^2/v_s^2$, and can be determined by requiring that R.H.S = 0 at 
some $r = r_{\scriptscriptstyle \rm F}$. We get,
\be
{\ln(1+r_{\scriptscriptstyle \rm F}/r_s) \over r_{\scriptscriptstyle \rm F}/r_s} = 1- \frac{v_\star^2}{v_s^2}
\ee
If the ratio $v_\star^2/v_s^2 \sim 0$ then $r_{\scriptscriptstyle \rm F}\sim 0$ from the above equation, which contradicts the assumption in deriving this solution as it is a supersonic solution and valid for $r>R$. 
It implies that the ratio needs to have a finite value so that we get,  $r_{\scriptscriptstyle \rm F} > R$. Therefore, in this case wind  starts with velocity
being equal to the sound speed at r = R and then stops at a distance $r_{\scriptscriptstyle \rm F}$. 

Another interesting result which can be 
deduced from the above equation is the condition for escape of the wind. If $v_\star = v_s$ then $r_{\scriptscriptstyle \rm F} = \infty$, which means 
that winds will escape the galaxy. Hence any heating occurring at the base need to somehow propel the gas at least 
to a speed of $\sim v_s$  so that the gas can escape the galaxy. 
\end{itemize}

\begin{figure}[h]
    \centering
    \includegraphics[scale=0.5]{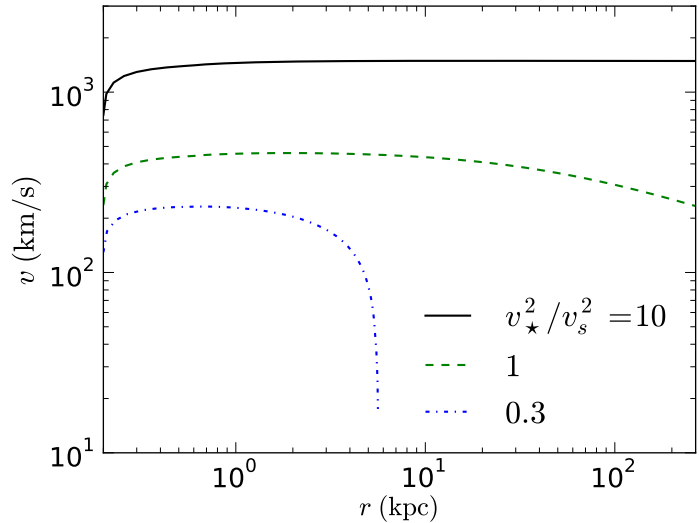}
    \caption{The wind velocity in the supersonic regime for the solution with Navarro-Frenk-White dark matter halo. The reference mass used is  $M_h = 10^{12} \Msun$. The solid, dashed and dotted line are for $v_\star^2/v_s^2=10, 1, 0.3$ respectively. The x-axis starts at $r=R=200$ pc and ends at the virial radius of the halo.
    }
    \label{fig:photo}
\end{figure}

In figure 1 we plot the supersonic wind velocity for a reference halo mass $M_h=10^{12}$ $M_\sun$ with other NFW parameters calculated using definitions in \S \ref{DMhalo} at $z=0$. The value of $v_s=236$ \kms for this halo. The solid line is for $v_\star^2=10v_s^2$ and in this case the injection is large so that the gravity of the halo does not affect the flow. That is why the solid line leaves the plot window and the wind velocity at largest distance is $\approx1500$ \kms which is double the value at $r=R$ in agreement with the result $v_\infty=2v_\star$ of previous section on the solution of CC85.  The dashed line represents the case 2 when $v_\star=v_s$ and the NFW gravity is important. Consequently the velocities are reduced and at the virial radius, we get a wind velocity $\sim 200$ \kms. The dotted line is for $v_\star^2 = 0.3v_s^2$ and it represents the case 3 discussed above. This line exhibits the flow which ends inside the galactic halo at a final radius, $r_{\scriptscriptstyle \rm F}\sim50$ kpc.  
We find that for the low $v_\star$ values, which implies a low value of velocity at the critical point, the gas stops at some point within the halo.
We note that similar solutions were obtained by \cite{Wang95} for power-law potential, and for specific values of
wind speed and mass loss rate, but the calculations were done numerically. The analytical model presented here
generalizes these solutions to NFW halo and to general values of wind speed. 

\section{Injection parameters and value of ${\MakeLowercase v}_\star$}
In the last section we discussed the importance of the ratio, 
$v_\star^2/v_s^2 = \dot E/(2 \dot M v_s^2)$. Here we would like to determine the possible values
of $v_\star$ which we will use for the rest of the analysis. 
We can define the  energy injection from supernovae  per unit time as
$
\dot E = \alpha f_{SN} \sfr\times (10^{51}  \ {\rm erg})
$
 where $10^{51}$ {\rm erg} is the energy output of a single supernova, $\alpha$ is the fraction of this energy retained by the gas after radiative energy losses, $\sfr$ is the SFR in solar mass per year and $f_{SN}$ stands for energy 
injection per year per solar mass of star formation. For a Kroupa-Chabrier IMF, $f_{SN}\sim 1.26\times 10^{-2}$.
Therefore we have,
\be
\dot E \approx \left[\alpha\ \sfr\ (4 \times 10^{41})\right]\ {\rm erg}\ {\rm s}^{-1}
\ee
The mass injection rate is written as
\be
\dot M  = \beta \mathcal{R}_f  \sfr
\ee
Where $\mathcal{R}_f$ is the return-fraction. Typically 30\% of the mass is returned to the ISM hence $\mathcal{R}_f\approx0.3$. The factor $\beta$ takes into account the entrainment of mass from ISM which can increase the overall mass injection. For the starburst galaxy M82, $\beta$ is in the range 1.0 to 2.8 which gives $\dot M \sim$ 1.4 \hbox{--} 3.6 $\Msun \ {\rm yr}^{-1}$ \citep{Strickland09}. \cite{Martin99} found that for a range of galaxies from low to high masses the mass loss rate roughly scales with SFR . We therefore use $\dot M\sim \sfr$, which corresponds to $\beta \mathcal{R}_f \sim 1$ .

With these value of $\dot E$ and $\dot M$ we estimate the value of $v_\star$ which is $(\dot E/2\dot M)^{1/2}=562\sqrt{\alpha}$ \kms. We 
consider two modes of energy injection from SNe. In the first case, almost 90\% of the energy of SNe is lost via radiation and only a small fraction goes into heating of the wind. For this 'quiescent mode', we use $\alpha = 0.1$ 
which gives $v_\star \approx 180$ km s$^{-1}$. 
In the other case, when the central injection region is dense and the supernovae are overlapping so that radiative losses reduce, and due to thermalization, 30\% to 100\% of the  supernova energy goes in heating the wind \citep{Strickland09}. This type of situation is evident in galaxies like M82 where the SFR is generally high and the injection regions are supposedly quite dense. For such a case $\alpha = 0.3\ \mbox{--}\ 1.0$ and we get $v_\star \approx 308\ \mbox{--}\ 562$ km s$^{-1}$. To represent this mode we take  $v_\star = 500$ \kms. We will call this the  'starburst mode'. 

The values of $v_\star$ chosen here brackets the range of the possible values.
We emphasize that this range of $v_\star$, when equated with sound speed corresponds to a  temperature range of roughly $\sim 0.2\ \hbox{--}\ 1$ keV,
 which is consistent with the hot wind temperatures observed in a wide range of galaxies \citep{Martin2005}. On the other hand the 
quiescent mode ($v_\star=180$ \kms) is suitable for the galaxies with low values of SFR, like our own Galaxy. This particular mode also yields interesting results, and we will discuss the implications in a later section.

\section{Winds in the presence of AGN}
The AGN is important as it gives a strong momentum injection to the gas via its radiation field. 
A large fraction of AGNs show the evidence of outflowing gas, and it is possible that all AGNs drive outflows
and they are observed when they are viewed edge-on. Theoretically, these outflows have been associated with the 
co-evolution of black holes and the bulge of the host galaxy \citep{Silk98,King03,King05}. 
How the AGN  interacts with  the ISM of the host galaxies and whether it can drive a large scale outflow which can escape the galaxy is an important question. AGN can affect the gas in the host galaxy indirectly where it produces fast nuclear winds which  shock the ISM into shells. The fate of these shells is then decided by the supply of energy and momentum injection from the inner regions. Apart from this indirect way AGN can also interact with the dust-rich ISM directly via its radiation field. This interaction is capable of driving large scale outflows \citep{MQT05}.      
Here we consider this mechanism and model the outflows as being driven
by (Eddington limited) continuum radiation from the black hole. 

\subsection{Effect of momentum injection from AGN}
 Momentum injection can be provided by the  AGN in several ways. Firstly it can be provided via the scattering of photons by the free-electrons. As the Thompson opacity is generally  small ($\kappa_T \approx 0.34$ cm$^2$ gm$^{-1}$ for a fully ionized gas with solar metallicity), this may be effective in the regions close to AGN where the densities 
are quite large and the radiation field  is strong. Another way  momentum is transferred to  produces outflows is
via line driving mechanism \citep{Murray95,Proga00}. We consider the  momentum injection via the absorption and scattering of AGN radiation by dust 
grains. Dust opacities are rather high and recent models of momentum driving of outflows due to AGNs and galactic radiation as well, consider the scattering of photons by dust grains. 
The gas is assumed to be coupled to these grains through momentum
coupling, and get dragged. We justify this assumption in Appendix E.

Let us derive the force due to momentum injection by AGN radiation ($f(r)$) which then will be substituted in equation \ref{central}. In its general form, the momentum injection from a radiation field  can be written as, 
\be
f(r) = \kappa\frac{\bf \mathcal{F}}{c} 
\ee
where $\kappa$ is the volume averaged opacity, ${\mathcal F}$ stands for the frequency integrated flux of radiation and c is the speed of light.
For spherical symmetry and for a  optically thin atmosphere, the radiation flux can be written as $\mathcal{F} = L/4\pi r^2$. 
Hence the force of radiation becomes $\kappa L/4\pi r^2c$. This force has an inverse square dependence on r, hence it can be represented as a factor ($\Gamma$) times the gravitational force of the black hole.  
Therefore, if the gravity of the central black hole is given by $f_{grav} = -G \mbh/r^2$ then in the presence of an outward radiation force,  the effective force is written as
$
f_{g,eff} = - {(1-\Gamma)G \mbh/ r^2 }
$ 
with 
$
\Gamma = \kappa L/(4\pi G \mbh c)
$.
For $\Gamma = 1$, the effective force is zero, and for the case of  Thompson scattering  the corresponding luminosity is called the Eddington luminosity ($L_E$). The luminosity required to exactly cancel the gravitational force may be different depending on the opacity and process responsible for momentum injection. For example, in case of scattering of UV light by dust, for which the opacity is roughly  3500 times the electron-scattering opacity \citep{Draine03}, only a luminosity of $\sim 0.001\ L_E$ is required to counter the black hole gravity.   However, it has been showed recently that most of the UV photon field from the AGN may get attenuated within a short distance because of the large optical depths in AGN environments \citep{Novak12}. In that case the re-radiated Infrared(IR) photons serve as the mainstay of AGN radiation (see also \cite{Dorod12}.
Therefore in this work we consider the momentum injection from IR radiation.  IR to dust scattering opacity is, $\kappa_{IR}=13$ cm$^2$ gm$^{-1}$, in K band \citep{Li01,Draine03} for a gas to dust ratio of 125. Opacity in IR is not as large as it is in UV, however we find that it is large enough to drive strong outflows in massive galaxies.
Hence the momentum injection force in our case becomes,
\be
f(r) = {\kappa_{IR} L\over4 \pi r^2 c} =  \Gamma{G\mbh \over r^2}
\ee
where $\Gamma = \kappa_{IR} L/(4\pi G \mbh c)$. We consider an Eddington limited AGN  ($L=L_{E}$), where $L_{E}= (4 \pi G \mbh c /\kappa_{T})$ is the usual Eddington luminosity for electron scattering.
This fixes the value of 
$
\Gamma={\kappa_{IR}/\kappa_{T}} \approx 38 \,
$,
for  our case.

We need to justify 
that the atmosphere is optically thin in IR so that we can work with a  constant $\Gamma$. In order to do this we estimate the optical depth for IR light.  Optical depth can be estimated as,  $\tau\sim\int\kappa\rho(r)dr$. 
Using  $\kappa_{IR}=13$ cm$^2$ gm$^{-1}$ and $\rho\sim10$ m$_{\rm p}$ cc$^{-1}$which, as we will see is an upper limit for density in our wind models, the optical depth comes out to be  $0.01$ at $20$ pc and $0.13$ at $200$ pc. Even at the edge of the injection region the optical depth is very small, hence we conclude that the atmosphere is optically thin to  IR radiation. Beyond $200$ pc, the density decreases rapidly, as the wind expands adiabatically in the supersonic regime. As we will see in next section, density becomes as low as $0.001$ m$_{\rm p}$ cc$^{-1}$ at $10$ kpc, hence the wind material stays optically thin to IR radiation at large distances as well. 


Let us  consider the supersonic section of the wind in which both SNe and AGN radiation are effective. (See Appendix B for the subsonic part of this wind.)
For the supersonic part the energy and the mass injection both are zero, although the gravity due to NFW halo and 
the effective force due to radiation and gravity from the central AGN are present. The total potential can be written as,
\be
\Phi_{total}(r) = \Phi_\bullet(r) +  \Phi_{\rm NFW}(r) =  -{G \mbh\over r} -2v_s^2\ \frac{\ln(1+r/r_s)}{r/r_s}
\ee 
With the aid of this total potential and the momentum injection term $f(r)=\Gamma{G\mbh/r^2}$,  the wind equation \ref{central} for supersonic part ($r>R$) can be written as,
\begin{eqnarray}
 \frac{\mach^2-1}{\mach^2 [(\gamma -1)\mach^2+2]}\frac{d\mach^2}{dr} =
 \frac{2}{r} - \frac{2}{\epsilon(r)} \left({(1-\Gamma) G\mbh\over r^2}+\frac{d\Phi_{NFW}(r)}{dr}\right)
 \label{agn3}
\end{eqnarray}
Next we integrate the energy equation directly as below,
\begin{eqnarray}
 && \left. \epsilon(r) - {(1-\Gamma) G\mbh\over r} + \Phi_{NFW}(r)\right|_R^r = 0
 \nonumber \\
\Rightarrow && \epsilon(r) = \left[2v_{\rm crit}^2 -2(1-\Gamma)\vbh^2 + \Phi_{\rm NFW}(R)\right]+ 2(1-\Gamma) \vbh^2{R\over r}  -\Phi_{\rm NFW}(r) 
\label{agn:4}
\end{eqnarray} 
where we have used $\epsilon(R) = 2 v_{\rm crit}^2= 2 v_\star^2-(1-\Gamma) \vbh^2$, from Appendix B for the subsonic section of this wind. We have also used  $\vbh^2 = G\mbh/2R$. The term inside the square bracket in the above equation is a constant, so if we substitute eq \ref{agn:4} in \ref{agn3} it becomes exactly integrable and we get,
\be
\ln\left|\delta_>(\mach)\right| = 2 \ln \left|r\right| + 2 \ln \left|2v_{\rm crit}^2 -2(1-\Gamma)\vbh^2 + \Phi_{\rm NFW}(R) + 2(1-\Gamma) \vbh^2{R\over r}   -\Phi_{\rm NFW}(r) \right|  + const.
\ee
Using the condition\footnote{By imposing this condition we have picked the solution which becomes supersonic at $r=R=200$ pc. To see the complete solution space the reader is referred to the $\mach$ versus r diagrams in Appendix \ref{lastapp}.}, $\delta_>(\mach) = 1 $ at $r=R$, and substituting the expression for NFW potential we get,
\be
\delta_>(\mach) = (r/R)^2 \left[1 - (1-\Gamma){\vbh^2\over v_{\rm crit}^2}\left(1-{R\over r} \right) -  {v_s^2\over  v_{\rm crit}^2} \left( \frac{\ln(1+R/r_s)}{R/r_s} - \frac{\ln(1+r/r_s)}{r/r_s}\right)\right]^2
\label{agnsuper}
\ee
where $v_{\rm crit}^2=  v_\star^2-(1-\Gamma) \vbh^2/2$. This equation gives the complete solution of the wind from a galaxy driven by
energy injection from SNe and  momentum injection from the AGN. Let us discuss some asymptotic behaviours,
\begin{itemize}
\item Terminal speed :\\
We use the Bernoulli equation \ref{agn:4} to obtain the terminal speed. Taking the limit $r\rightarrow\infty$ in equation \ref{agn:4} and neglecting the sound speed  we obtain the following general expression for terminal speed of SNe and AGN driven wind from a NFW halo,
\be
v_\infty = 2 \left(v_\star^2 + \frac{3}{2}(\Gamma-1)\vbh^2 -  v_s^2\right)^{1/2}
\ee
Here $v_s^2=GM_h/(2\mathcal{C}r_s)$.  In the absence of dark matter halo and the AGN we can neglect $v_s$ and $\vbh$ which gives $v_\infty = 2v_\star$. 
However in presence of NFW gravity but no AGN we get the  relation $v_\infty =  2\sqrt{v_\star^2-v_s^2}$. 
For dwarf galaxies, the effect of NFW gravity can be neglected, and therefore
the wind speed in starburst mode is expected to be $v_\infty \approx 1000$ 
\kms, consistent with observations of winds in starburst galaxy \object{M82}.
If the black hole is  massive so that the $\vbh$ dominates then by neglecting $v_\star$ and $v_s$ and putting $\Gamma=38$,  we get $v_\infty\approx15\vbh$, which for a black hole of mass $\sim10^9$ $\Msun$ estimates to $v_\infty\sim1500$ \kms.  

\item Behaviour with r : \\
In equation \ref{agnsuper} if the SNe injection term ($v_\star$) is dominant then we get $\mach^3\propto r^{2}$. When the wind is moving with constant terminal speed,  $T\propto r^{-4/3}$ and $\rho\propto r^{-2}$ as derived in \S 2.1. If the AGN provides the main driving then neglecting $v_s$ and $v_\star$ from equation \ref{agnsuper},
for  large r, we once again retain $\mach^3\propto r^2$, and hence the scaling $T\propto r^{-4/3}$. The density scales as $\rho\propto r^{-2}$.
\end{itemize}

It is clear that, apart from the $v_\star$, the wind properties depend on two more terms. One involves $\vbh$ which is a parameter of the  black hole and the other being $v_s$, which depends on the mass of NFW halo. To proceed further we need to learn whether the  black hole mass is somehow related to dark matter halo.

Observations show that the black hole mass is related to the  velocity dispersion of the central spheroidal bulge component  of the galaxies, and the relation can be written as
\be
\log \left({\mbh \over \Msun}\right) = a + b \log\left({\sigma \over 200\ {\rm km\ s^{-1}}}\right)
\ee
 This correlation has been extensively studied in the literature, with slightly different values for $a$ and $b$. We will list a few of them.   \cite{Gebhardt00} was one of the earliest to report the correlation with $a\simeq8.08\pm0.07$ and $b=3.75\pm0.3$. \cite{Ferrarese02} gives $a\simeq8.22\pm0.08$ and $b=4.58\pm0.52$. Values in \cite{Tremaine02} reads $a=8.13\pm0.06$ and $b=4.02\pm0.32$. For our analysis we use $a=8.12$ and $b = 4.24$ from a recent study by  \cite{Gultekin09}.

We also use a relation between the $\sigma$ and circular speed $\vvir$.
\cite{Ferrarese02} reported a correlation between the circular speed in outskirts of galaxy and the $\sigma$, given roughly as $\vvir\propto\sigma^{0.84\pm0.09}$. Similar relations were also deduced by \cite{Baes03} and \cite{Pizzella05}.  All these relation are very close to the linear relation  $\vvir=\sqrt{2}\sigma$ for spherical mass distribution \citep{BT08}, found in massive ellipsoidal galaxies where the bulge to total mass ratio is unity (see also \citealt{Volonteri11}). Hence for our analysis we use $v_c = \sqrt{2}\sigma$. 

The relation between $v_c$ and $\sigma$ breaks down for low mass galaxies \citep{Ferrarese02}. 
It is easy to understand this as the lower mass galaxies may admit a bulge to total mass ratio less than unity. Hence the galaxies with smaller bulges will have the mass of their black holes smaller than the one expected from $\mbh\hbox{--}\sigma\hbox{--}v_c$ relation. On similar grounds, \cite{Kormendy11} concluded against the co-evolution of central black hole and dark matter halo. However,  they also showed that a cosmic conspiracy causes $v_c$ to correlate with $\sigma$ for massive galaxies ($\vvir\gtrsim200$\kms).

We would like to emphasize here that even if the  $v_c\hbox{--} \sigma$ relation breaks down below a $v_c\lesssim 200$ \kms,  still it does not make a difference in our results because the AGN term is effective only for very massive systems. In equation \ref{agn3} it is the relative values of $\vbh$ and $v_s$, which govern the dynamics.  For $\vvir$=200 \kms, using $c=12$ we get $v_s = (GM_h/2\mathcal{C}r_s)^{1/2} \approx 400$ \kms which is  larger than $(\sqrt{\Gamma-1})\vbh \approx 100$ \kms, and the black hole term is even smaller for $\vvir < 200$ \kms.

Using the $\mbh\hbox{--}\sigma$ relation provided by \cite{Gultekin09}, aided with the relation $\sigma=v_c/\sqrt{2}$, where $v_c$ is given by equation \ref{defvc} we arrive at the following relation between  black hole mass and the halo mass (see also \cite{Volonteri11}).
\be
\left(\mbh \over 0.74 \times 10^7\ \msun \right)^{1/\sqrt{2}} \simeq  \ \left({M_h\over 10^{12}\ \Msun}\right)\left[{\Omega_m\over\Omega_m^z}{\Delta_c(z)\over 18\pi^2}\right]^{1/2}(1+\zvir)^{3/2}
\label{BHDM}
\ee
where $\zvir$ is the redshift at which the halo collapsed and got viriallized.
This relation can be used to determine $\vbh$ as a function of halo mass through, $\vbh = (G\mbh/2R)^{1/2}$.  Substituting $\vbh$ and $v_s=\sqrt{GM_h/2r_s\mathcal{C}}$ in  equation \ref{agnsuper} with $r_s$ and $\mathcal{C}$ determined from the redshift dependent definitions for NFW parameters given in \S \ref{DMhalo}. This enables us to compute the mach numbers as a function of $r$ for the galaxy of any desired halo mass($M_h$) and its halo collapsing at any desired redshift. The mach numbers can then be converted to the velocity simply by multiplying it with the sound speed which can be obtained using the relation $c_s^2 = \epsilon(r)/(\mach^2/2+1/(\gamma-1))$.

\begin{figure*}[h]
    \centering
    \includegraphics[scale=0.7]{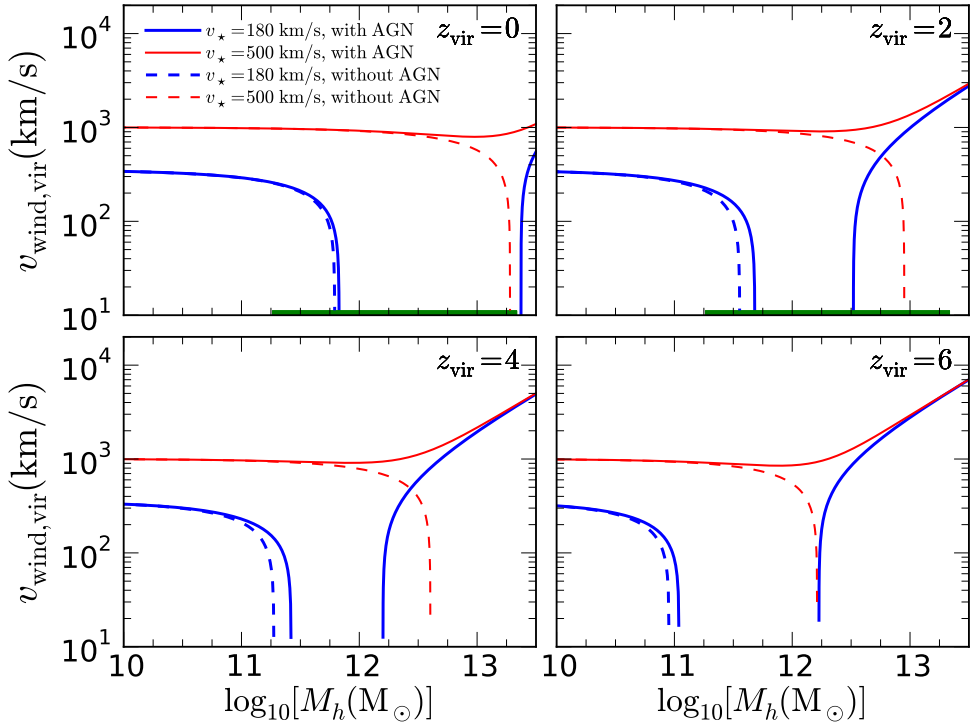}
    \caption{Wind velocity at virial radius as a function of halo mass for four different redshifts of collapse, $\zvir=0,2,4,6$ in the panels from top-left to bottom-right  respectively. The thin red lines in each panel are for starburst mode ($v_\star=500$ \kms) and thick blue ones are for quiescent mode ($v_\star=180$ \kms). Dashed line represent the outflow speeds without AGN. The green bar on x-axis in top two panels gives the range of galaxies in which the gas reservoirs in  halos are observed \citep{Tumlinson12}. 
    }
    \label{fig:final}
\end{figure*}

In figure \ref{fig:final} we plot the wind velocities at virial radius\footnote{Wind velocity at virial radius ($v_{\rm wind,vir}$) need not be equal to the the terminal speed at infinity ($v_\infty$) as the later is calculated by using the fact that the gravitational force at the infinity is zero while in case of $v_{\rm wind,vir}$ there may be a contribution from NFW gravity at virial radius.} 
($v_{\rm wind,vir}$) as a function of halo mass($M_h$).   The velocities are obtained by solving equation (\ref{agnsuper}) with inputs from equation \ref{BHDM} and definitions in section \ref{DMhalo}. We show the results corresponding to four different value of $\zvir$ in four panels. red lines are for the starburst mode $v_\star=500$ \kms and the blue lines are for the quiescent mode  $v_\star=180$ \kms. The dashed lines represent the solution without AGN and these show that the wind velocities cut of at some halo mass as the gravity becomes strong enough to counter the energy injection. For larger value of $v_\star$ this cut-off occurs at a larger halo mass.

Consider a situation in which for low halo masses the AGN driving term  is smaller than the NFW gravity term.  
This implies that the  wind velocity at virial radius in low mass halos decreases with increase in halo mass. However the black hole mass also increases with the halo mass.  Since the slope of $\mbh\hbox{--} M_h$ relation is greater than 
unity, hence the rate of  increase of the black hole or AGN term is larger and at a  particular halo mass it overcomes the NFW  gravity term. Hence for largest galaxies we should see an increase in wind velocity with halo mass, which indeed is the case as  shown by rising  solid  lines for massive halos. One can further compare the  solid  lines with  dashed  lines where the latter represent the case without the AGN and does not show any winds in high mass galaxies, which is a confirmation of the fact that in high mass galaxies outflows are driven by AGN.

The thick solid blue curve is special as it features the wind quenching due to NFW gravity as well as the high velocity winds due to AGN.  There is a falling part of the curve which exhibits the cut off at some halo mass and then there is another part which rises at some larger halo mass. Hence there exist a range of halo masses  roughly within $10^{11\hbox{--}13}\Msun$,  which do not have escaping winds. The exact value of this range depends on the  redshift of collapse ($\zvir$). For example at $\zvir=0$ the  rising part of the thick blue curve, which shows the effect of AGN, starts rising beyond a halo mass of $10^{13}\Msun$ while for $\zvir=2$ it rises roughly at a halo mass $\sim10^{12}\Msun$. This is easy to understand as the AGN driving depends on the black hole mass which does increase with redshift (see equation \ref{BHDM}). Also the falling part of the thick blue curve ends at a smaller halo mass for a larger value of $\zvir$, because the value of $v_s$ increases with redshift. We would like to mention here that the recent detection  of gas reservoirs in the halos of galaxies by \cite{Tumlinson12} covers roughly the similar range in halo masses shown by the green bar on x-axis in upper two panels of figure \ref{fig:final}.

If we focus on the wind velocities in lower to intermediate mass halos $M_h<10^{12}\Msun$, we find that AGN never dominate in these  and if there are winds they have to be driven by starbursts and SNe. We note that the wind velocities in low mass galaxies fall in the range $400\hbox{--}1000$ \kms, depending on the efficiency  of the energy injection process. These velocities agree with the ones inferred from the X-ray temperatures of the superwind regions in dwarf straburst and luminous infrared galaxies \citep{Heckman00, Martin99}. However in case of  galaxies with halo mass, $M_h\gtrsim10^{12.5}\Msun$ the wind velocities either exceed 1000 \kms or they are quenched, depending on whether the AGN is present or not.

We note that in low mass galaxies where the outflows are driven by SNe, the wind velocity, $v_{\rm wind}\lesssim1000$ \kms. However this limit is exceeded when an AGN is present, since the curves with AGN show  wind velocities  $\gtrsim 1000$ \kms. This  reveals the presence of a dividing line of 1000 \kms between SNe and AGN domination in velocity space as well. 

In a hot wind with velocity of $v_{\rm wind}$ the neutral clumps in the wind can be dragged via the ram pressure. Maximum velocity these clouds can achieve is the velocity of the hot wind. As mentioned above, $v_{\rm wind}$ is always within 1000 \kms at the low mass end where  SNe injection dominates, therefore the cold clouds should also be outflowing with velocity lower than 1000 \kms. On the other hand at the higher mass end, where the AGN dominates, the velocities may exceed this limit.   
Interestingly, the observed  outflow speeds of the neutral clouds in diverse galaxies like dwarf starbursts,  ULIRGs, post-starburst galaxies and Low-ionization BAL quasars also follow this trend \citep{Tremonti07,Sturm11,Trump06}.
 
It has been debated in the literature that the neutral cold/warm  winds in ULIRGs are driven by  ram pressure of the hot wind and/or radiation from stars in the galaxy or by the AGN. If we consider the winds driven by stellar radiation then the wind speed is roughly 3 times the circular speed of the galaxy \citep{MQT05,Martin2005,Sharmaetal12}. For a massive ULIRG with a circular speed $\sim 300$ \kms the wind velocity predicted by radiation driven wind model will be $900$ \kms.  On the other hand, if ram pressure is the driving mechanism, then also the velocities of the continuous hot wind and hence of the neutral clouds can not exceed 1000 \kms unless an AGN is present. 
Therefore we emphasize that the observations of wind velocity in excess of 1000 \kms indicate  the presence of an AGN.

\subsection{Wind properties with distance : Implications for gas observed in galactic halos}
In the last section we have established that for a particular mass range the galactic winds may not escape the galaxy. Therefore these galaxies are not very important for the intergalactic medium (IGM) enrichment. Interestingly our Milky Way with a total mass roughly $\sim  10^{12}\Msun$, also falls in this mass range. However an important question arises for these type of galaxies, as to whether or not these galaxies can retain all the gas in the disk even though they can contain the gas inside $\rvir$.
\begin{figure*}[h]
    \centering
    \includegraphics[scale=0.7]{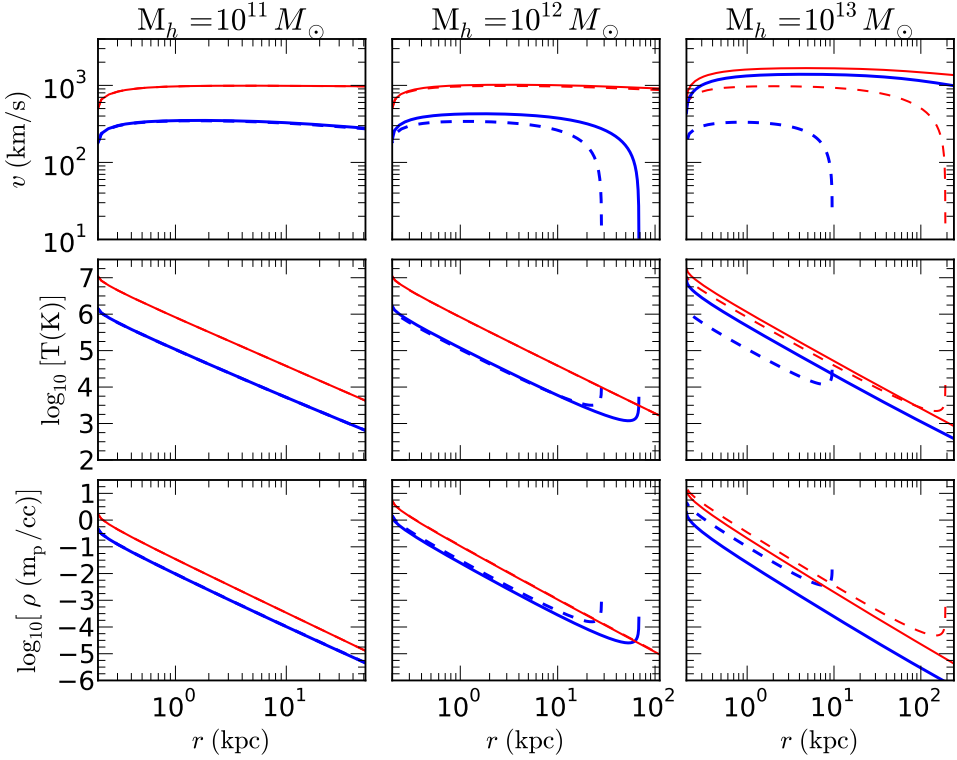}
    \caption{The wind properties as a function of distance from the base. Legend is same as in figure \ref{fig:final}. From left to right, the three columns correspond to three different halo masses, representing  a dwarf galaxy, a Milky Way size galaxy and a giant galaxy  respectively. The x-axis in all the plots extend upto the corresponding virial radius. The solid lines represent the winds in presence of an Eddington limited AGN and the dashed lines, the case of no AGN. Thin red lines represent the starburst mode ($v_\star =500$ \kms) and the thick blue lines are for the quiescent mode ($v_\star=180$ \kms). The middle row contains the temperature profiles of winds in three galaxies. The lowermost row shows the densities. To calculate  densities we have used $\sfr=1,3,10$ $\Msun$ yr$^{-1}$ for $v_\star=180$ \kms, and $\sfr=10,30,100$ $\Msun$ for $v_\star=500$ \kms, for three galaxies respectively.
    }
    \label{fig:agn_mw}
\end{figure*}

We show in this section that infact these galaxies have outflows which spill their gas reservoir throughout the halo. In figure \ref{fig:agn_mw} we show the radial dependence of wind properties for three galaxies which differ in the values of their halo mass. We have considered the halos collapsing at a redshift of $\zvir=2$ which corresponds to a look-back time of roughly $\sim 10$ {\rm Gyr}. We choose this collapse redshift, as a fiducial value, in order to model galaxies like Milky Way and more massive galaxies,
whose halos were already in place by $z \sim 2$. The thick disk in our Galaxy shows that Milky Way underwent its last major merger before $\sim 10$ Gyr, which corresponds to
$z\sim 2$ \citep{Gilmore02}. Therefore our results can be compared with observations of winds at low redshift ($z \le 2$) universe.

In figure \ref{fig:agn_mw}, the thick blue lines refer to the quiescent mode ($v_\star=180$ \kms) and the thin red lines represent the starburst mode ($v_\star=500 $ \kms). 
The solid lines denote the effect of AGN activity and dashed lines
consider only the SNe injection. The upper panels plot the velocity as a function of radius, and the
middle and bottom panels plot the temperature and densities respectively. The densities are estimated using the relation $\rho=\dot M/4\pi v r^2$, where $\dot M \sim \sfr$. For three galaxies with $M_h = 10^{11},10^{12},10^{13}\Msun$, we have used $\sfr=1,3,10$ $\Msun$yr$^{-1}$  for quiescent mode and $\sfr=10,30,100$ $\Msun$yr$^{-1}$ for starburst mode respectively.

The curves show that for low mass galaxies, all types of  winds (with or without AGN, quiescent and starburst mode) escape the virial radius. At the other extreme, for massive galaxies, winds with AGN activity can escape, but without an AGN, they stop at a distance within the halo ($\sim 10\hbox{--}200$ kpc). 
The gas temperature and density slightly rises at this final halting point due to adiabatic compression. We would like to mention here that the radiative cooling can be important for these particular cases as the steady solution ceases to exist beyond a point. Even if the cooling time is shorter than the flow time initially, after  many flow crossing times the cooling will become effective  which may cause thermal instability. This can lead to formation of clouds whose fate then will be decided by the physical properties in their environment (see also \citealt{Wang95}).

For the intermediate mass galaxy ($M_h\sim 10^{12}\Msun$), we find that both of the thick blue lines (i.e., quiescent mode of star formation with or without AGN) are contained inside the virial radius. This implies that the wind needs  strong starburst activity in order to escape the galaxy irrespective of whether or not AGN is present. The dashed thick blue line corresponds to a quiescent mode of star formation without AGN and roughly corresponds to our own Galaxy. Interestingly, we find that a slow wind is possible, which  may extend to a distance of $\sim 20$ kpc. This can explain the recent observations of clouds  roughly at $10\hbox{--}20$ kpc in our halo \citep{Keeney06}. 

The solutions which end inside  $\rvir$ are important in the wake of recent observation of warm-hot gas clouds  in the halo of our galaxy  and for other galaxies as well \citep{Tumlinson12}. Also recent simulations confirm these gas reservoirs around the galaxies of intermediate masses.  
Recent works find that although in the general scenario of galaxy formation the intermediate mass galaxies are efficient in retaining their baryons, these galaxies do not retain {\it all} of the baryons. It appears that only 20\% to 30\% of baryons are converted to stars in these galaxies as well \citep{Somerville06,Moster10,Behroozi10}. Our results provide a natural explanation for the missing baryons in these intermediate mass galaxies.

\subsection{Cosmological implications}
Here we derive a relation between the stellar mass and the halo mass using the results for SNe and AGN driven outflows in the previous section. In the the scenario of hierarchical structure formation low mass galaxies form at earlier times and post formation history is influenced by mergers and the periods of enhanced star formation activity. The  semi-analytical modeling (SAM)(see \citealt{Baugh06}), which uses simple recipes for feedback and  follows the structure formation according to $\Lambda$CDM model, can  explain the observed properties of galaxies \citep{Somerville06}.
 In recent years there have been a growing amount of observational evidence that the massive black holes were already in place at high redshifts. Also, it is well accepted that the massive galaxies formed their stars at earlier epochs as they appear redder at present times, compared to the younger galaxies in which star formation is still going on \citep[and references therein]{Fontanot09}. This phenomenon, which are commonly referred as 'downsizing', indicate a possible role the massive black holes would have played in quenching the star formation in massive galaxies at earlier epochs \citep{Somerville04,Granato04,Croton06}.
We follow the semi-analytical scheme proposed in \cite{Granato04} for which an approximate analysis is provided in the appendix A of \cite{Shankar06}, using which,
 we can write the rate of change of cool gas in halos as,
\be
\dot M_{cold}(t) =  \frac{M_{infall}(t)}{t_c} - \sfr(t) + \mathcal{R}_f \sfr(t) - \mathcal{L} \sfr(t) \,,
\label{eq_infall}
\ee
where $\mathcal{R}_f$ is the
return fraction of stars whose value is $0.3$ for a Salpeter IMF. $\mathcal{L} \sfr$ is the mass loss rate, $t_c$ is the cooling time. $M_{infall}(t)$ is the mass available in the halo for infall at a time $t$  . The Equation \ref{eq_infall}  can be solved to obtain a time independant solution in large time limit and
yields a stellar mass at present epoch as (see eqn 17 in \citealt{Shankar06}),
\be
M_{\star} = f_{surv} \frac{f_{cosm}M_h}{1 - \mathcal{R}_f + \mathcal{L}} \,,
\label{mstar}
\ee
where $f_{cosm}=1/6$ is the cosmic baryon ratio, $f_{surv}$ is the fraction of stars surviving up to now and its value is approximately  $0.6$ for a Salpeter IMF.
As $f_{surv},\ f_{cosm}$ and $\mathcal{R}_f$ are constants, therefore we get $M_{star} \propto M_h/\mathcal{L}$.
In case of SNe feedback the feedback factor is written as $\mathcal{L}=\frac{\alpha f_{SN} (10^{51} \, {\rm erg})}{E_{bind}}$, where $E_{bind}=[\vvir^2 f(c)(1+f_{cosm})/2]$ with $f(c)$ as given in \cite{Mo98},  represent the binding energy of gas in the halo per unit mass. The quantity in the numerator is nothing but  $\dot E/\dot M = 2v_\star^2$ where $v_\star$ is the value of velocity at the critical point for the case of only SNe injection and no AGN.  When the AGN is also present we can use the modified value of velocity at the critical point given in equation \ref{agn:4} which is $v_{\rm crit}^2 = v_\star^2-(1-\Gamma)\vbh^2/2$. Hence in the case of both SNe and AGN feedback we can write the loss term $\mathcal{L}$ as, 
\be
\mathcal{L} =\frac{2 v_{\rm crit}^2}{{1 \over 2}\vvir^2 f(c) (1+f_{cosm})} = \frac{2[v_\star^2 - 
({1-\Gamma\over2})\vbh^2]}{{1 \over 2}\vvir^2 f(c) (1+f_{cosm})} 
\label{eq_Loss}
\ee 
$v_{\rm crit}$ is  the velocity at the critical point and $\vvir$ the virial velocity of the halo. 
While the velocity at the critical point measures the strength of SNe and/or AGN as a mechanism of mass expulsion,  the velocity at the virial radius is a measure of binding energy of the halo.

For NFW halo, $\vvir\propto M_h^{1/3}$. Also $\vbh^2 \propto \mbh \propto M_h^{1.41}$ (see equation \ref{BHDM}).
Using these we get
\be
M_{\star} \propto \frac{M_h}{1-\mathcal{R}_f + {4 v_\star^2 + C M_h^{1.41}\over D M_h^{2/3}}} \propto \frac{M_h^{5/3}}{A +  C M_h^{1.41}}
\ee
where $C$, $D$ are constants and $A=[0.8 f(c)v_c^2+4v_\star^2]$. From the above relation we find that for small halo masses we have $M_{\star} \propto M_h^{5/3}$ and for large halo masses, when the AGN 
dominates, we get $M_{\star} \propto M_h^{0.26}$.  The break occurs and $M_\star/M_h$ peaks roughly in the range $\sim 10^{12\hbox{--}12.5}\Msun$ as the ratio $v_{\rm crit}^2/\vvir^2$ becomes minimum around this mass. 
\begin{figure*}[h]
    \centering
    \includegraphics[scale=0.7]{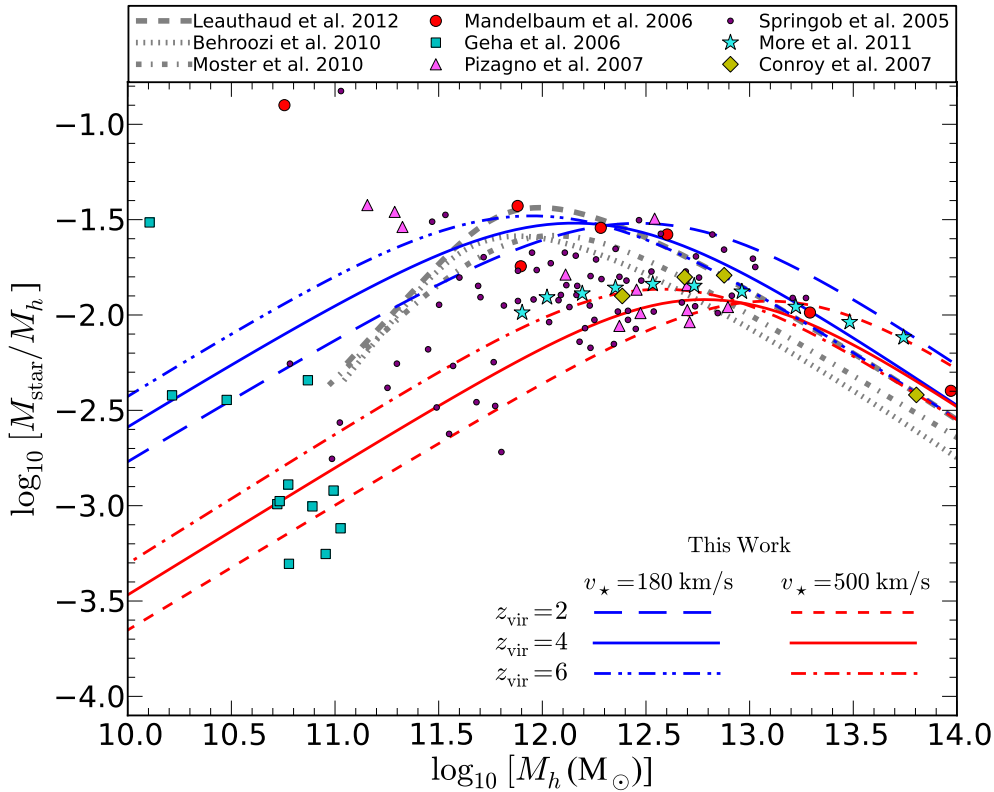}
    \caption{SHMR at present time from theoretical considerations in this work is compared with the observational data for individual galaxies courtesy  Alexie Leauthaud. The upper three (blue) lines represent the quiescent mode ($v_\star=180$ \kms)  and the lower three (red) lines represent the starburst mode ($v_\star=500$ \kms) of galaxies. In these the dashed, solid and dash-dotted lines correspond to collapse redshifts, $\zvir=2$, 4 and 6 respectively. The  data points are from \citealt{Mandelbaum06} : filled red circles, \citealt{Geha06}:filled green squares, \citealt{Pizagno07} : filled purple triangles, \citealt{Springob05} : dots, \citealt{More11} : stars, \citealt{Conroy07} : diamonds.  The thick gray dashed, dotted and dash-dotted lines represent the results of halo abundance matching from \cite{Leauthaud12}, \cite{Behroozi10} and \cite{Moster10} respectively. 
    }
    \label{final_2}
 \end{figure*}

 In Figure \ref{final_2}, we plot the SHMR at present day, against the halo mass  obtained by using equation \ref{mstar}, \ref{eq_Loss} and \ref{BHDM}.   The set of upper three blue lines correspond to quiescent mode ($v_\star=180$ \kms) and the lower three red line represent the starburst mode ($v_\star=500$ \kms). In each set the dashed, solid and the dash-dotted lines correspond to three different redshifts of collapse, $\zvir=2$, 4 and $6$ respectively. We stress here that the lines denote the value of SHMR at present day ($z=0$) for the galaxies which got viriallized at a particular $\zvir$.

The squares, dots,  and triangles  in Figure 4 represent the stellar and halo mass  data inferred from works on Tully fisher relation by \cite{Geha06}, \cite{Springob05} and \cite{Pizagno07} respectively. The stars and diamonds represent the data from \citep{More11} and \citep{Conroy07} who deduced the stellar and halo mass using  stellar dynamics. Red circles shows the estimates based on weak lensing by \citep{Mandelbaum06}. For details on the  data sets the reader is referred to \cite{Leauthaud12} and \cite{Blanton08} and the original papers for the the data sets.
We have also shown the SHMR obtained by the technique of halo abundance matching \citep{Moster10,Behroozi10,Leauthaud12}  using broken thick gray lines. At the higher mass end we note that our results agree with these works. To be more specific we find a stellar to halo mass slope of $0.26$ as mentioned above, which is in agreement with the value $0.29$ deduced in \cite{Behroozi10}. 
The  slope at higher mass end depends  on the $\mbh-M_h$ relation, which is still being debated in the literature. However, even if we use a different scaling like $\mbh \propto \sigma^{4.02}$ given by \cite{Tremaine02}, then the slope at high mass end becomes $0.33$ which is also in agreement with observations and other works. 

The slope at low mass end as found by halo abundance matching  is roughly
$\sim 2.2$, which is larger than $5/3$ from theoretical considerations in this work and others (e.g. \citealt{DekelWoo03}). however,  it is possible that other physical processes not considered in our model
can explain the discrepancy. For example, if $v_\star$ depends on halo mass, with the efficiency of SNe energy
injection being larger in low mass halos, the slopes can be reconciled. 

If one naively compares the lines from our analytical calculation with the observational data points for individual galaxies then the plot seems to say something interesting. By looking at the data for low to intermediate mass galaxies \citep{Geha06,Springob05,Pizagno07} one may infer that the data points lie systematically below the results of halo abundance matching. There are a lot of galaxies which have lower stellar content than expected from halo abundance matching. However, if we compare these with our lower three red lines which are for starburst mode then these data points can be reconciled. As we have already described, the outflow activity in low to intermediate mass galaxies is governed  by SNe and starbursts. The two modes we are following in this paper represent two extreme efficiencies of energy injection by SNe. Quiescent galaxies like ours lie on one extreme and the  violent galaxy like M82 on the other extreme.  All the galaxies in low to intermediate halo mass range ($\lesssim10^{12.5}\Msun$) are covered  between these two extremes. Interestingly in Figure 4 also, most of the data points lie between the blue and red lines of our theoretical model which represent the quiescent and starburst mode of energy injection respectively. This is not mere a coincidence given the simplicity of our model and shows the importance of outflows in shaping the galaxies. 

In the above analysis, we have assumed that the relation given in equation \ref{mstar}, which is valid in the long time limit, gives the stellar mass in the present day universe. 
In the  hierarchical structure formation scenario where small halos form first and undergo periods of high merger rates, this assumption may not be completely valid. 
However, recent observations and theoretical works have shown that cosmic downsizing mitigates the effects of hierarchical structure formation models, and that 
massive galaxies are believed to form stars at high redshift after which they evolve passively. In this regard, the consistency of our results with observations of massive galaxies becomes important.

\section{Discussions}
In this paper we have studied galactic outflows driven by SNe injection and AGN radiation. We have calculated the outflow properties in halos ranging from low to high mass. 
The treatment is analytical which enabled us to extract fundamental results for these feedback processes. Here we recapitulate the results and the caveats.

We have found  that the NFW gravity causes the outflows to stop inside virial radius in intermediate to high mass halos.   In Appendix C, we have  shown the solutions for winds
from NFW halo in terms of Mach number, where the closed contours show the importance of the halo. In other words, gravitational force of the halo causes the flow to stop inside the virial radius  if the energy injection from SNe is not large. We note that a minimum value of SNe energy injection so as to produce $v_\star \sim v_s$ is required for the gas to escape the galaxy. Our models in which the gas can not escape the galaxy, provides a natural explanation for  the circumgalactic material observed inside the halos and also predict mild winds in quiescent star forming galaxies such as our Galaxy.

When an AGN component is included, the contours in the Mach number$\hbox{--}$radius plot can open up for
massive galaxies ($M_h\gtrsim 10^{12.5}\Msun$). This implies that AGN radiation can become important in winds from massive galaxies (such as ULIGs), reaching a wind velocity of  $v_{\rm wind}\gtrsim 1000$ \kms.  If we consider that neutral clouds are entrained in this wind then the speeds of the cold clouds can be at the most equal to $v_{\rm wind}$. This result is  
consistent with observed outflow speeds in post-starburst galaxies, ULIRGs and Low-ionization quasars at intermediate to high redshifts \citep[e.g.][]{Tremonti07,Sturm11}. We have derived a general expression for the terminal speed of the wind from the NFW halo which can be written as $v_\infty = (4v_\star^2+6(\Gamma-1)\vbh^2-4v_s^2)^{1/2}$, where $v_\star$ is the contribution from SNe and the term with $\vbh$ stands for the momentum injection from AGN. According to our results 1000 \kms is an upper limit for the starburst driven winds while in case of AGN driving which is possible in high mass galaxies only, the  velocities always exceed 1000 \kms. We note here that this limit holds for large scale escaping outflows for all galaxies; however there may be exceptional cases where the system is going through a period of extreme  star formation, and even without AGN, the wind speed at a few kpc can be more than 1000 \kms  \citep{Alex12}. These episodic winds are likely to be driven by the radiation from galaxy  as the galaxy becomes highly luminous due to extreme star formation \citep{Sharmaetal12}.

We have also shown that our results can explain the observed trends of stellar to halo mass ratio. We find that the stellar mass scales with the halo mass as $M_\star\propto M_h^{0.26}$ at the higher mass end and $M_\star\propto M_h^{1.67}$ at the lower mass end. The slope at higher mass end agrees with the observations and abundance matching results. We find that the large scatter in observational data at the lower mass end is due to the fact that the efficiency of energy injection is different in different galaxies. 
We would like to mention here that we have used a simple recipe to calculate the stellar mass corresponding to a particular halo mass using the outputs from our wind models. Implementing feedback recipes from our wind models into  a full semi analytical work is beyond the scope of this paper. Also we have assumed spherical symmetry for our calculation which is justified considering the large length scale ($\sim 100$ kpc)  of these outflows. However in the vicinity of the disk the effect of gravity due to a flattened system may be important as explored using a self-similar model in \citet{Bardeen78}. We leave the problem of finding streamlines in a cylindrical geometry  for these outflows to a future work.

We note that the reprocessed IR radiation is sufficient to drive strong outflows from high mass  galaxies and the resulting feedback is enough to explain the mismatch between 
stellar mass and halo mass at high mass end. This is important as one need not to rely  on UV radiation which is supposed to be attenuated quickly within a small distance from the center.

The injection of energy and mass in our model occurs in the central region. However recent observations show the evidence of outflows emerging from individual star clusters 
which may be situated away from the center of the galaxies \citep[e.g.][]{Schwartz06}. It becomes however analytically complicated to take into account the contribution from these clusters and 
combine it with the effects of a central AGN. It is however clear that effects of the feedback would be higher if mass and energy injection occurs even beyond the central region and from a large number of clusters.

The coupling between gas and dust is shown to be mediated via momentum coupling. However one may question the survival of dust grains. For this we refer to the  observational evidence for dust  in the spectra of AGNs, which shows that the dust does survive and it may do so by residing in 
small clumps around AGN as discussed in \cite{Krolik88}.

We have considered momentum transfer from AGN radiation in optically thin limit. In actual practice the situation can be quite complicated and it may not be exactly accurate to use a constant $\Gamma$. We hope that the simulations of winds with full radiation transfer covering from small  to large length scales will be able to verify the simple ideas presented here.
We have used a relation between the black hole mass and the dark matter halo mass. We have projected this relation backward in time using scalings of NFW parameters with 
redshift. There are observational evidence that black hole masses at high redshifts are generally higher, however it is hard to predict the correct scaling of black hole mass with redshift as the formation and growth of black holes is a complicated problem in itself.     

We have not considered the effects of radiative cooling in the present work. Radiative cooling is important for the thermodynamics of the outflow. The CC85 solution with implementation of cooling has been studied in the past \citep{Silich04,Tenorio07,Wunsch11,Silich11}. These calculations conclude that cooling (if not catastrophic) does not affect the velocity and density but causes the temperature to decay more rapidly. Wind solutions with cooling can not be worked out analytically for general values of parameters and one has to use the numerical computation, for particular cases. 
We have discussed the hot phase of the outflows which is tractable using analytic hydrodynamics. However in actual practice the physics of hot, cold and molecular phase of galactic outflows might be tangled to each other . The study of all these components demands state of the art numerical simulations. We hope to study the multi-phase character of these winds in a future paper. 

To summarize, we have found analytical solutions for  SNe and AGN driven winds from realistic dark matter halo. Our results show that the two feedback processes operate effectively at two ends of the galaxy luminosity function as expected. The wind velocities for escaping winds resulting from our calculations explains a variety of observations. We find that AGN can drive the gas to speeds $\gtrsim1000$ \kms. We find an intermediate mass range in which the outflows can be highly suppressed and for these halo masses the gas can not escape into the IGM. However, the gas is still driven to large distances within the halo. This result explains the recent observations of gas reservoirs in our Galaxy and other galaxies. Using the results of our analytical models, we have derived a stellar mass to halo mass relation using simple recipes.  We find the derived $M_\star\hbox{--}M_h$ relation matches the results of observations and halo abundance matching.  Thus,  our findings  provide a possible explanation for missing baryons in galaxies. 

We thank  Mitchell Begelman for critical reading of the manuscript.   We are grateful to Alexie Leauthaud for providing the observational data used in Figure 4.  We thank Tim Heckman, Rachel Somerville and Prateek Sharma  for fruitful discussions. We also thank an anonymous  referee for constructive comments.

\bibliography{ref}

\begin{thebibliography}{123}
\expandafter\ifx\csname natexlab\endcsname\relax\def\natexlab#1{#1}\fi

\bibitem[{{Alexander} {et~al.}(2010){Alexander}, {Swinbank}, {Smail},
  {McDermid}, \& {Nesvadba}}]{Alexander10}
{Alexander}, D.~M., {Swinbank}, A.~M., {Smail}, I., {McDermid}, R., \&
  {Nesvadba}, N.~P.~H. 2010, \mnras, 402, 2211

\bibitem[{{Baes} {et~al.}(2003){Baes}, {Buyle}, {Hau}, \& {Dejonghe}}]{Baes03}
{Baes}, M., {Buyle}, P., {Hau}, G.~K.~T., \& {Dejonghe}, H. 2003, \mnras, 341,
  L44

\bibitem[{{Bardeen} \& {Berger}(1978)}]{Bardeen78}
{Bardeen}, J.~M., \& {Berger}, B.~K. 1978, \apj, 221, 105

\bibitem[{{Baugh}(2006)}]{Baugh06}
{Baugh}, C.~M. 2006, Reports on Progress in Physics, 69, 3101

\bibitem[{{Behroozi} {et~al.}(2010){Behroozi}, {Conroy}, \&
  {Wechsler}}]{Behroozi10}
{Behroozi}, P.~S., {Conroy}, C., \& {Wechsler}, R.~H. 2010, \apj, 717, 379

\bibitem[{{Binney}(2004)}]{Binney04}
{Binney}, J. 2004, \mnras, 347, 1093

\bibitem[{{Binney} \& {Tremaine}(2008)}]{BT08}
{Binney}, J., \& {Tremaine}, S. 2008, {Galactic Dynamics: Second Edition}
  (Princeton University Press)

\bibitem[{{Blanton} {et~al.}(2008){Blanton}, {Geha}, \& {West}}]{Blanton08}
{Blanton}, M.~R., {Geha}, M., \& {West}, A.~A. 2008, \apj, 682, 861

\bibitem[{{Bouch{\'e}} {et~al.}(2012){Bouch{\'e}}, {Hohensee}, {Vargas},
  {Kacprzak}, {Martin}, {Cooke}, \& {Churchill}}]{Bouche12}
{Bouch{\'e}}, N., {Hohensee}, W., {Vargas}, R., {et~al.} 2012, \mnras, 3207

\bibitem[{{Bower} {et~al.}(2006){Bower}, {Benson}, {Malbon}, {Helly}, {Frenk},
  {Baugh}, {Cole}, \& {Lacey}}]{Bower06}
{Bower}, R.~G., {Benson}, A.~J., {Malbon}, R., {et~al.} 2006, \mnras, 370, 645

\bibitem[{{Breitschwerdt} {et~al.}(1991){Breitschwerdt}, {McKenzie}, \&
  {V{\"o}lk}}]{Breitschwerdt1991}
{Breitschwerdt}, D., {McKenzie}, J.~F., \& {V{\"o}lk}, H.~J. 1991, \aap, 245,
  79

\bibitem[{{Bryan} \& {Norman}(1998)}]{Bryan98}
{Bryan}, G.~L., \& {Norman}, M.~L. 1998, \apj, 495, 80

\bibitem[{{Bullock} {et~al.}(2001){Bullock}, {Kolatt}, {Sigad}, {Somerville},
  {Kravtsov}, {Klypin}, {Primack}, \& {Dekel}}]{Bullock01}
{Bullock}, J.~S., {Kolatt}, T.~S., {Sigad}, Y., {et~al.} 2001, \mnras, 321, 559

\bibitem[{{Burke}(1968)}]{Burke1968}
{Burke}, J.~A. 1968, \mnras, 140, 241

\bibitem[{{Cattaneo} {et~al.}(2006){Cattaneo}, {Dekel}, {Devriendt},
  {Guiderdoni}, \& {Blaizot}}]{Cattaneo06}
{Cattaneo}, A., {Dekel}, A., {Devriendt}, J., {Guiderdoni}, B., \& {Blaizot},
  J. 2006, \mnras, 370, 1651

\bibitem[{{Chattopadhyay} {et~al.}(2012){Chattopadhyay}, {Sharma}, {Nath}, \&
  {Ryu}}]{Chatto12}
{Chattopadhyay}, I., {Sharma}, M., {Nath}, B.~B., \& {Ryu}, D. 2012, \mnras,
  423, 2153

\bibitem[{{Chevalier} \& {Clegg}(1985)}]{CC85}
{Chevalier}, R.~A., \& {Clegg}, A.~W. 1985, \nat, 317, 44

\bibitem[{{Conroy} {et~al.}(2007){Conroy}, {Prada}, {Newman}, {Croton}, {Coil},
  {Conselice}, {Cooper}, {Davis}, {Faber}, {Gerke}, {Guhathakurta}, {Klypin},
  {Koo}, \& {Yan}}]{Conroy07}
{Conroy}, C., {Prada}, F., {Newman}, J.~A., {et~al.} 2007, \apj, 654, 153

\bibitem[{{Cooper} {et~al.}(2008){Cooper}, {Bicknell}, {Sutherland}, \&
  {Bland-Hawthorn}}]{Cooper08}
{Cooper}, J.~L., {Bicknell}, G.~V., {Sutherland}, R.~S., \& {Bland-Hawthorn},
  J. 2008, \apj, 674, 157

\bibitem[{{Cooper} {et~al.}(2009){Cooper}, {Bicknell}, {Sutherland}, \&
  {Bland-Hawthorn}}]{Cooper09}
---. 2009, \apj, 703, 330

\bibitem[{{Croton} {et~al.}(2006){Croton}, {Springel}, {White}, {De Lucia},
  {Frenk}, {Gao}, {Jenkins}, {Kauffmann}, {Navarro}, \& {Yoshida}}]{Croton06}
{Croton}, D.~J., {Springel}, V., {White}, S.~D.~M., {et~al.} 2006, \mnras, 365,
  11

\bibitem[{{Debuhr} {et~al.}(2012){Debuhr}, {Quataert}, \& {Ma}}]{Debuhr12}
{Debuhr}, J., {Quataert}, E., \& {Ma}, C.-P. 2012, \mnras, 420, 2221

\bibitem[{Dekel \& Silk(1986)}]{Dekel86}
Dekel, A., \& Silk, J. 1986, \apj, 303, 39

\bibitem[{{Dekel} \& {Woo}(2003)}]{DekelWoo03}
{Dekel}, A., \& {Woo}, J. 2003, \mnras, 344, 1131

\bibitem[{{Di Matteo} {et~al.}(2005){Di Matteo}, {Springel}, \&
  {Hernquist}}]{Matteo05}
{Di Matteo}, T., {Springel}, V., \& {Hernquist}, L. 2005, \nat, 433, 604

\bibitem[{{Diamond-Stanic} {et~al.}(2012){Diamond-Stanic}, {Moustakas},
  {Tremonti}, {Coil}, {Hickox}, {Robaina}, {Rudnick}, \& {Sell}}]{Alex12}
{Diamond-Stanic}, A.~M., {Moustakas}, J., {Tremonti}, C.~A., {et~al.} 2012,
  \apjl, 755, L26

\bibitem[{{Dorodnitsyn} {et~al.}(2011){Dorodnitsyn}, {Bisnovatyi-Kogan}, \&
  {Kallman}}]{Dorod12}
{Dorodnitsyn}, A., {Bisnovatyi-Kogan}, G.~S., \& {Kallman}, T. 2011, \apj, 741,
  29

\bibitem[{{Draine}(2003)}]{Draine03}
{Draine}, B.~T. 2003, \araa, 41, 241

\bibitem[{{Dunn} {et~al.}(2010){Dunn}, {Bautista}, {Arav}, {Moe}, {Korista},
  {Costantini}, {Benn}, {Ellison}, \& {Edmonds}}]{Dunn10}
{Dunn}, J.~P., {Bautista}, M., {Arav}, N., {et~al.} 2010, \apj, 709, 611

\bibitem[{{Everett} \& {Murray}(2007)}]{Everett07}
{Everett}, J.~E., \& {Murray}, N. 2007, \apj, 656, 93

\bibitem[{{Faucher-Gigu{\`e}re} \& {Quataert}(2012)}]{Faucher12}
{Faucher-Gigu{\`e}re}, C.-A., \& {Quataert}, E. 2012, \mnras, 3450

\bibitem[{{Ferrarese}(2002)}]{Ferrarese02}
{Ferrarese}, L. 2002, \apj, 578, 90

\bibitem[{{Feruglio} {et~al.}(2010){Feruglio}, {Maiolino}, {Piconcelli},
  {Menci}, {Aussel}, {Lamastra}, \& {Fiore}}]{Feruglio10}
{Feruglio}, C., {Maiolino}, R., {Piconcelli}, E., {et~al.} 2010, \aap, 518,
  L155

\bibitem[{{Fontanot} {et~al.}(2009){Fontanot}, {De Lucia}, {Monaco},
  {Somerville}, \& {Santini}}]{Fontanot09}
{Fontanot}, F., {De Lucia}, G., {Monaco}, P., {Somerville}, R.~S., \&
  {Santini}, P. 2009, \mnras, 397, 1776

\bibitem[{{Fu} \& {Stockton}(2009)}]{Fu09}
{Fu}, H., \& {Stockton}, A. 2009, \apj, 690, 953

\bibitem[{{Gebhardt} {et~al.}(2000){Gebhardt}, {Bender}, {Bower}, {Dressler},
  {Faber}, {Filippenko}, {Green}, {Grillmair}, {Ho}, {Kormendy}, {Lauer},
  {Magorrian}, {Pinkney}, {Richstone}, \& {Tremaine}}]{Gebhardt00}
{Gebhardt}, K., {Bender}, R., {Bower}, G., {et~al.} 2000, \apjl, 539, L13

\bibitem[{{Geha} {et~al.}(2006){Geha}, {Blanton}, {Masjedi}, \&
  {West}}]{Geha06}
{Geha}, M., {Blanton}, M.~R., {Masjedi}, M., \& {West}, A.~A. 2006, \apj, 653,
  240

\bibitem[{{Gilman}(1972)}]{Gilman72}
{Gilman}, R.~C. 1972, \apj, 178, 423

\bibitem[{{Gilmore} {et~al.}(2002){Gilmore}, {Wyse}, \& {Norris}}]{Gilmore02}
{Gilmore}, G., {Wyse}, R.~F.~G., \& {Norris}, J.~E. 2002, \apjl, 574, L39

\bibitem[{{Granato} {et~al.}(2004){Granato}, {De Zotti}, {Silva}, {Bressan}, \&
  {Danese}}]{Granato04}
{Granato}, G.~L., {De Zotti}, G., {Silva}, L., {Bressan}, A., \& {Danese}, L.
  2004, \apj, 600, 580

\bibitem[{{Grimes} {et~al.}(2009){Grimes}, {Heckman}, {Aloisi}, {Calzetti},
  {Leitherer}, {Martin}, {Meurer}, {Sembach}, \& {Strickland}}]{Grimes09}
{Grimes}, J.~P., {Heckman}, T., {Aloisi}, A., {et~al.} 2009, \apjs, 181, 272

\bibitem[{{G{\"u}ltekin} {et~al.}(2009){G{\"u}ltekin}, {Richstone}, {Gebhardt},
  {Lauer}, {Tremaine}, {Aller}, {Bender}, {Dressler}, {Faber}, {Filippenko},
  {Green}, {Ho}, {Kormendy}, {Magorrian}, {Pinkney}, \& {Siopis}}]{Gultekin09}
{G{\"u}ltekin}, K., {Richstone}, D.~O., {Gebhardt}, K., {et~al.} 2009, \apj,
  698, 198

\bibitem[{{Heckman} {et~al.}(1993){Heckman}, {Lehnert}, \& {Armus}}]{Heckman93}
{Heckman}, T.~M., {Lehnert}, M.~D., \& {Armus}, L. 1993, in Astrophysics and
  Space Science Library, Vol. 188, The Environment and Evolution of Galaxies,
  ed. J.~M. {Shull} \& H.~A. {Thronson}, 455

\bibitem[{{Heckman} {et~al.}(2000){Heckman}, {Lehnert}, {Strickland}, \&
  {Armus}}]{Heckman00}
{Heckman}, T.~M., {Lehnert}, M.~D., {Strickland}, D.~K., \& {Armus}, L. 2000,
  \apjs, 129, 493

\bibitem[{{Holzer} \& {Axford}(1970)}]{Holzer70}
{Holzer}, T.~E., \& {Axford}, W.~I. 1970, \araa, 8, 31

\bibitem[{{Hopkins} {et~al.}(2012){Hopkins}, {Quataert}, \&
  {Murray}}]{Hopkins12}
{Hopkins}, P.~F., {Quataert}, E., \& {Murray}, N. 2012, \mnras, 421, 3522

\bibitem[{Ipavich(1975)}]{Ipavich1975}
Ipavich, F.~M. 1975, \apj, 196, 107

\bibitem[{{Johnson} \& {Axford}(1971)}]{Johnson71}
{Johnson}, H.~E., \& {Axford}, W.~I. 1971, \apj, 165, 381

\bibitem[{{Keeney} {et~al.}(2006){Keeney}, {Danforth}, {Stocke}, {Penton},
  {Shull}, \& {Sembach}}]{Keeney06}
{Keeney}, B.~A., {Danforth}, C.~W., {Stocke}, J.~T., {et~al.} 2006, \apj, 646,
  951

\bibitem[{{King}(2003)}]{King03}
{King}, A. 2003, \apjl, 596, L27

\bibitem[{{King}(2005)}]{King05}
---. 2005, \apjl, 635, L121

\bibitem[{{King} {et~al.}(2011){King}, {Zubovas}, \& {Power}}]{King11}
{King}, A.~R., {Zubovas}, K., \& {Power}, C. 2011, \mnras, 415, L6

\bibitem[{{Komatsu} {et~al.}(2009){Komatsu}, {Dunkley}, {Nolta}, {Bennett},
  {Gold}, {Hinshaw}, {Jarosik}, {Larson}, {Limon}, {Page}, {Spergel},
  {Halpern}, {Hill}, {Kogut}, {Meyer}, {Tucker}, {Weiland}, {Wollack}, \&
  {Wright}}]{Komatsu09}
{Komatsu}, E., {Dunkley}, J., {Nolta}, M.~R., {et~al.} 2009, \apjs, 180, 330

\bibitem[{{Kormendy} \& {Bender}(2011)}]{Kormendy11}
{Kormendy}, J., \& {Bender}, R. 2011, \nat, 469, 377

\bibitem[{{Kornei} {et~al.}(2012){Kornei}, {Shapley}, {Martin}, {Coil}, {Lotz},
  {Schiminovich}, {Bundy}, \& {Noeske}}]{Kornei12}
{Kornei}, K.~A., {Shapley}, A.~E., {Martin}, C.~L., {et~al.} 2012, ArXiv
  e-prints

\bibitem[{{Krolik} \& {Begelman}(1988)}]{Krolik88}
{Krolik}, J.~H., \& {Begelman}, M.~C. 1988, \apj, 329, 702

\bibitem[{{Kurosawa} \& {Proga}(2009)}]{Kurosawa09}
{Kurosawa}, R., \& {Proga}, D. 2009, \apj, 693, 1929

\bibitem[{{Larson}(1974)}]{larson74}
{Larson}, R.~B. 1974, \mnras, 169, 229

\bibitem[{{Leauthaud} {et~al.}(2012){Leauthaud}, {Tinker}, {Bundy}, {Behroozi},
  {Massey}, {Rhodes}, {George}, {Kneib}, {Benson}, {Wechsler}, {Busha},
  {Capak}, {Cort{\^e}s}, {Ilbert}, {Koekemoer}, {Le F{\`e}vre}, {Lilly},
  {McCracken}, {Salvato}, {Schrabback}, {Scoville}, {Smith}, \&
  {Taylor}}]{Leauthaud12}
{Leauthaud}, A., {Tinker}, J., {Bundy}, K., {et~al.} 2012, \apj, 744, 159

\bibitem[{{Li} \& {Draine}(2001)}]{Li01}
{Li}, A., \& {Draine}, B.~T. 2001, \apj, 554, 778

\bibitem[{{Mandelbaum} {et~al.}(2006){Mandelbaum}, {Seljak}, {Kauffmann},
  {Hirata}, \& {Brinkmann}}]{Mandelbaum06}
{Mandelbaum}, R., {Seljak}, U., {Kauffmann}, G., {Hirata}, C.~M., \&
  {Brinkmann}, J. 2006, \mnras, 368, 715

\bibitem[{{Marcolini} {et~al.}(2005){Marcolini}, {Strickland}, {D'Ercole},
  {Heckman}, \& {Hoopes}}]{Marcolini05}
{Marcolini}, A., {Strickland}, D.~K., {D'Ercole}, A., {Heckman}, T.~M., \&
  {Hoopes}, C.~G. 2005, \mnras, 362, 626

\bibitem[{{Martin}(1999)}]{Martin99}
{Martin}, C.~L. 1999, \apj, 513, 156

\bibitem[{{Martin}(2005)}]{Martin2005}
---. 2005, \apj, 621, 227

\bibitem[{{Mathews} \& {Baker}(1971)}]{Mathews71}
{Mathews}, W.~G., \& {Baker}, J.~C. 1971, \apj, 170, 241

\bibitem[{{McQuillin} \& {McLaughlin}(2012)}]{McQ12}
{McQuillin}, R.~C., \& {McLaughlin}, D.~E. 2012, \mnras, 423, 2162

\bibitem[{Mo {et~al.}(1998)Mo, Mao, \& White}]{Mo98}
Mo, H.~J., Mao, S., \& White, S. D.~M. 1998, \mnras, 295, 319

\bibitem[{{More} {et~al.}(2011){More}, {van den Bosch}, {Cacciato}, {Skibba},
  {Mo}, \& {Yang}}]{More11}
{More}, S., {van den Bosch}, F.~C., {Cacciato}, M., {et~al.} 2011, \mnras, 410,
  210

\bibitem[{{Morganti} {et~al.}(2007){Morganti}, {Holt}, {Saripalli},
  {Oosterloo}, \& {Tadhunter}}]{Morganti07}
{Morganti}, R., {Holt}, J., {Saripalli}, L., {Oosterloo}, T.~A., \&
  {Tadhunter}, C.~N. 2007, \aap, 476, 735

\bibitem[{{Moster} {et~al.}(2010){Moster}, {Somerville}, {Maulbetsch}, {van den
  Bosch}, {Macci{\`o}}, {Naab}, \& {Oser}}]{Moster10}
{Moster}, B.~P., {Somerville}, R.~S., {Maulbetsch}, C., {et~al.} 2010, \apj,
  710, 903

\bibitem[{{Mu{\~n}oz-Cuartas} {et~al.}(2011){Mu{\~n}oz-Cuartas}, {Macci{\`o}},
  {Gottl{\"o}ber}, \& {Dutton}}]{Munoz11}
{Mu{\~n}oz-Cuartas}, J.~C., {Macci{\`o}}, A.~V., {Gottl{\"o}ber}, S., \&
  {Dutton}, A.~A. 2011, \mnras, 411, 584

\bibitem[{{Murray} {et~al.}(1995){Murray}, {Chiang}, {Grossman}, \&
  {Voit}}]{Murray95}
{Murray}, N., {Chiang}, J., {Grossman}, S.~A., \& {Voit}, G.~M. 1995, \apj,
  451, 498

\bibitem[{{Murray} {et~al.}(2011){Murray}, {M{\'e}nard}, \&
  {Thompson}}]{Murray11}
{Murray}, N., {M{\'e}nard}, B., \& {Thompson}, T.~A. 2011, \apj, 735, 66

\bibitem[{{Murray} {et~al.}(2005){Murray}, {Quataert}, \& {Thompson}}]{MQT05}
{Murray}, N., {Quataert}, E., \& {Thompson}, T.~A. 2005, \apj, 618, 569

\bibitem[{{Nath} \& {Silk}(2009)}]{NathSilk2009}
{Nath}, B.~B., \& {Silk}, J. 2009, \mnras, 396, L90

\bibitem[{{Navarro} {et~al.}(1996){Navarro}, {Frenk}, \& {White}}]{NFW96}
{Navarro}, J.~F., {Frenk}, C.~S., \& {White}, S.~D.~M. 1996, \apj, 462, 563

\bibitem[{{Novak} {et~al.}(2012){Novak}, {Ostriker}, \& {Ciotti}}]{Novak12}
{Novak}, G.~S., {Ostriker}, J.~P., \& {Ciotti}, L. 2012, ArXiv e-prints:
  1203.6062

\bibitem[{{Oppenheimer} \& {Dav{\'e}}(2006)}]{Oppenheimer06}
{Oppenheimer}, B.~D., \& {Dav{\'e}}, R. 2006, \mnras, 373, 1265

\bibitem[{{Parker}(1965)}]{Parker65}
{Parker}, E.~N. 1965, \ssr, 4, 666

\bibitem[{{Pizagno} {et~al.}(2007){Pizagno}, {Prada}, {Weinberg}, {Rix},
  {Pogge}, {Grebel}, {Harbeck}, {Blanton}, {Brinkmann}, \& {Gunn}}]{Pizagno07}
{Pizagno}, J., {Prada}, F., {Weinberg}, D.~H., {et~al.} 2007, \aj, 134, 945

\bibitem[{{Pizzella} {et~al.}(2005){Pizzella}, {Corsini}, {Dalla Bont{\`a}},
  {Sarzi}, {Coccato}, \& {Bertola}}]{Pizzella05}
{Pizzella}, A., {Corsini}, E.~M., {Dalla Bont{\`a}}, E., {et~al.} 2005, \apj,
  631, 785

\bibitem[{{Proga} {et~al.}(2000){Proga}, {Stone}, \& {Kallman}}]{Proga00}
{Proga}, D., {Stone}, J.~M., \& {Kallman}, T.~R. 2000, \apj, 543, 686

\bibitem[{{Puchwein} \& {Springel}(2012)}]{Puchwein12}
{Puchwein}, E., \& {Springel}, V. 2012, ArXiv e-prints: 1205.2694

\bibitem[{{Risaliti} \& {Elvis}(2010)}]{Risaliti10}
{Risaliti}, G., \& {Elvis}, M. 2010, \aap, 516, A89

\bibitem[{{Rupke} {et~al.}(2005{\natexlab{a}}){Rupke}, {Veilleux}, \&
  {Sanders}}]{RupkeAGN05}
{Rupke}, D.~S., {Veilleux}, S., \& {Sanders}, D.~B. 2005{\natexlab{a}}, \apj,
  632, 751

\bibitem[{{Rupke} {et~al.}(2005{\natexlab{b}}){Rupke}, {Veilleux}, \&
  {Sanders}}]{Rupke05}
---. 2005{\natexlab{b}}, \apjs, 160, 115

\bibitem[{{Rupke} \& {Veilleux}(2011)}]{Rupke11}
{Rupke}, D.~S.~N., \& {Veilleux}, S. 2011, \apjl, 729, L27

\bibitem[{Samui {et~al.}(2008)Samui, Subramanian, \& Srianand}]{Samui2008}
Samui, S., Subramanian, K., \& Srianand, R. 2008, \mnras, 385, 783

\bibitem[{{Schwartz} {et~al.}(2006){Schwartz}, {Martin}, {Chandar},
  {Leitherer}, {Heckman}, \& {Oey}}]{Schwartz06}
{Schwartz}, C.~M., {Martin}, C.~L., {Chandar}, R., {et~al.} 2006, \apj, 646,
  858

\bibitem[{{Shankar} {et~al.}(2006){Shankar}, {Lapi}, {Salucci}, {De Zotti}, \&
  {Danese}}]{Shankar06}
{Shankar}, F., {Lapi}, A., {Salucci}, P., {De Zotti}, G., \& {Danese}, L. 2006,
  \apj, 643, 14

\bibitem[{{Sharma} \& {Nath}(2012)}]{SharmaNath12}
{Sharma}, M., \& {Nath}, B.~B. 2012, \apj, 750, 55

\bibitem[{{Sharma} {et~al.}(2011){Sharma}, {Nath}, \&
  {Shchekinov}}]{Sharmaetal12}
{Sharma}, M., {Nath}, B.~B., \& {Shchekinov}, Y. 2011, \apjl, 736, L27

\bibitem[{{Silich} {et~al.}(2011){Silich}, {Bisnovatyi-Kogan}, {Tenorio-Tagle},
  \& {Mart{\'{\i}}nez-Gonz{\'a}lez}}]{Silich11}
{Silich}, S., {Bisnovatyi-Kogan}, G., {Tenorio-Tagle}, G., \&
  {Mart{\'{\i}}nez-Gonz{\'a}lez}, S. 2011, \apj, 743, 120

\bibitem[{{Silich} {et~al.}(2003){Silich}, {Tenorio-Tagle}, \&
  {Mu{\~n}oz-Tu{\~n}{\'o}n}}]{Silich03}
{Silich}, S., {Tenorio-Tagle}, G., \& {Mu{\~n}oz-Tu{\~n}{\'o}n}, C. 2003, \apj,
  590, 791

\bibitem[{{Silich} {et~al.}(2004){Silich}, {Tenorio-Tagle}, \&
  {Rodr{\'{\i}}guez-Gonz{\'a}lez}}]{Silich04}
{Silich}, S., {Tenorio-Tagle}, G., \& {Rodr{\'{\i}}guez-Gonz{\'a}lez}, A. 2004,
  \apj, 610, 226

\bibitem[{{Silk} \& {Nusser}(2010)}]{Silk10}
{Silk}, J., \& {Nusser}, A. 2010, \apj, 725, 556

\bibitem[{{Silk} \& {Rees}(1998)}]{Silk98}
{Silk}, J., \& {Rees}, M.~J. 1998, \aap, 331, L1

\bibitem[{{Simis} \& {Woitke}(2004)}]{Simis03}
{Simis}, Y., \& {Woitke}, P. 2004, in Asymptotic giant branch stars, ed. H.~J.
  {Habing} \& H.~{Olofsson}, Astronomy and Astrophysics Library, Springer, 291

\bibitem[{{Somerville} {et~al.}(2008){Somerville}, {Hopkins}, {Cox},
  {Robertson}, \& {Hernquist}}]{Somerville06}
{Somerville}, R.~S., {Hopkins}, P.~F., {Cox}, T.~J., {Robertson}, B.~E., \&
  {Hernquist}, L. 2008, \mnras, 391, 481

\bibitem[{{Somerville} {et~al.}(2004){Somerville}, {Moustakas}, {Mobasher},
  {Gardner}, {Cimatti}, {Conselice}, {Daddi}, {Dahlen}, {Dickinson},
  {Eisenhardt}, {Lotz}, {Papovich}, {Renzini}, \& {Stern}}]{Somerville04}
{Somerville}, R.~S., {Moustakas}, L.~A., {Mobasher}, B., {et~al.} 2004, \apjl,
  600, L135

\bibitem[{{Springel} {et~al.}(2005){Springel}, {Di Matteo}, \&
  {Hernquist}}]{Springel05}
{Springel}, V., {Di Matteo}, T., \& {Hernquist}, L. 2005, \apjl, 620, L79

\bibitem[{{Springob} {et~al.}(2005){Springob}, {Haynes}, {Giovanelli}, \&
  {Kent}}]{Springob05}
{Springob}, C.~M., {Haynes}, M.~P., {Giovanelli}, R., \& {Kent}, B.~R. 2005,
  \apjs, 160, 149

\bibitem[{{Stinson} {et~al.}(2012){Stinson}, {Brook}, {Prochaska}, {Hennawi},
  {Shen}, {Wadsley}, {Pontzen}, {Couchman}, {Quinn}, {Macci{\`o}}, \&
  {Gibson}}]{Stinson12}
{Stinson}, G.~S., {Brook}, C., {Prochaska}, J.~X., {et~al.} 2012, \mnras, 3506

\bibitem[{{Strickland} \& {Heckman}(2009)}]{Strickland09}
{Strickland}, D.~K., \& {Heckman}, T.~M. 2009, \apj, 697, 2030

\bibitem[{{Strickland} {et~al.}(2004){Strickland}, {Heckman}, {Colbert},
  {Hoopes}, \& {Weaver}}]{Strickland04}
{Strickland}, D.~K., {Heckman}, T.~M., {Colbert}, E.~J.~M., {Hoopes}, C.~G., \&
  {Weaver}, K.~A. 2004, \apj, 606, 829

\bibitem[{{Strickland} \& {Stevens}(2000)}]{Strickland00}
{Strickland}, D.~K., \& {Stevens}, I.~R. 2000, \mnras, 314, 511

\bibitem[{{Sturm} {et~al.}(2011){Sturm}, {Gonz{\'a}lez-Alfonso}, {Veilleux},
  {Fischer}, {Graci{\'a}-Carpio}, {Hailey-Dunsheath}, {Contursi}, {Poglitsch},
  {Sternberg}, {Davies}, {Genzel}, {Lutz}, {Tacconi}, {Verma}, {Maiolino}, \&
  {de Jong}}]{Sturm11}
{Sturm}, E., {Gonz{\'a}lez-Alfonso}, E., {Veilleux}, S., {et~al.} 2011, \apjl,
  733, L16

\bibitem[{{Suchkov} {et~al.}(1994){Suchkov}, {Balsara}, {Heckman}, \&
  {Leitherer}}]{Suchkov94}
{Suchkov}, A.~A., {Balsara}, D.~S., {Heckman}, T.~M., \& {Leitherer}, C. 1994,
  \apj, 430, 511

\bibitem[{{Sutherland} \& {Dopita}(1993)}]{Suth93}
{Sutherland}, R.~S., \& {Dopita}, M.~A. 1993, \apjs, 88, 253

\bibitem[{{Tenorio-Tagle} {et~al.}(2007){Tenorio-Tagle}, {W{\"u}nsch},
  {Silich}, \& {Palou{\v s}}}]{Tenorio07}
{Tenorio-Tagle}, G., {W{\"u}nsch}, R., {Silich}, S., \& {Palou{\v s}}, J. 2007,
  \apj, 658, 1196

\bibitem[{{Tremaine} {et~al.}(2002){Tremaine}, {Gebhardt}, {Bender}, {Bower},
  {Dressler}, {Faber}, {Filippenko}, {Green}, {Grillmair}, {Ho}, {Kormendy},
  {Lauer}, {Magorrian}, {Pinkney}, \& {Richstone}}]{Tremaine02}
{Tremaine}, S., {Gebhardt}, K., {Bender}, R., {et~al.} 2002, \apj, 574, 740

\bibitem[{{Tremonti} {et~al.}(2007){Tremonti}, {Moustakas}, \&
  {Diamond-Stanic}}]{Tremonti07}
{Tremonti}, C.~A., {Moustakas}, J., \& {Diamond-Stanic}, A.~M. 2007, \apjl,
  663, L77

\bibitem[{{Trump} {et~al.}(2006){Trump}, {Hall}, {Reichard}, {Richards},
  {Schneider}, {Vanden Berk}, {Knapp}, {Anderson}, {Fan}, {Brinkman},
  {Kleinman}, \& {Nitta}}]{Trump06}
{Trump}, J.~R., {Hall}, P.~B., {Reichard}, T.~A., {et~al.} 2006, \apjs, 165, 1

\bibitem[{{Tumlinson} {et~al.}(2011){Tumlinson}, {Thom}, {Werk}, {Prochaska},
  {Tripp}, {Weinberg}, {Peeples}, {O'Meara}, {Oppenheimer}, {Meiring}, {Katz},
  {Dav{\'e}}, {Ford}, \& {Sembach}}]{Tumlinson12}
{Tumlinson}, J., {Thom}, C., {Werk}, J.~K., {et~al.} 2011, Science, 334, 948

\bibitem[{{Uhlig} {et~al.}(2012){Uhlig}, {Pfrommer}, {Sharma}, {Nath},
  {En{\ss}lin}, \& {Springel}}]{Uhlig12}
{Uhlig}, M., {Pfrommer}, C., {Sharma}, M., {et~al.} 2012, \mnras, 423, 2374

\bibitem[{{van de Voort} \& {Schaye}(2012)}]{Vandevoort12}
{van de Voort}, F., \& {Schaye}, J. 2012, ArXiv e-prints: 1207.5512

\bibitem[{{Veilleux} {et~al.}(2005){Veilleux}, {Cecil}, \&
  {Bland-Hawthorn}}]{Veilleux05}
{Veilleux}, S., {Cecil}, G., \& {Bland-Hawthorn}, J. 2005, \araa, 43, 769

\bibitem[{{Villar-Mart{\'{\i}}n} {et~al.}(2011){Villar-Mart{\'{\i}}n},
  {Humphrey}, {Delgado}, {Colina}, \& {Arribas}}]{Villar11}
{Villar-Mart{\'{\i}}n}, M., {Humphrey}, A., {Delgado}, R.~G., {Colina}, L., \&
  {Arribas}, S. 2011, \mnras, 418, 2032

\bibitem[{{Volonteri} \& {Stark}(2011)}]{Volonteri11}
{Volonteri}, M., \& {Stark}, D.~P. 2011, \mnras, 417, 2085

\bibitem[{{Wang}(1995)}]{Wang95}
{Wang}, B. 1995, \apj, 444, 590

\bibitem[{{Westmoquette} {et~al.}(2012){Westmoquette}, {Clements}, {Bendo}, \&
  {Khan}}]{West12}
{Westmoquette}, M.~S., {Clements}, D.~L., {Bendo}, G.~J., \& {Khan}, S.~A.
  2012, \mnras, 424, 416

\bibitem[{{W{\"u}nsch} {et~al.}(2011){W{\"u}nsch}, {Silich}, {Palou{\v s}},
  {Tenorio-Tagle}, \& {Mu{\~n}oz-Tu{\~n}{\'o}n}}]{Wunsch11}
{W{\"u}nsch}, R., {Silich}, S., {Palou{\v s}}, J., {Tenorio-Tagle}, G., \&
  {Mu{\~n}oz-Tu{\~n}{\'o}n}, C. 2011, \apj, 740, 75

\bibitem[{{Wyithe} \& {Loeb}(2003)}]{Wyithe03}
{Wyithe}, J.~S.~B., \& {Loeb}, A. 2003, \apj, 595, 614

\end{thebibliography}



\appendix

\section{Appendix A: Detailed derivation of wind equation}
The equations (1) and (2)  can be written as 
\be
{2\over r} + {1\over \rho}{d\rho\over dr} + {1\over v}\dvr = {\dot{m}\over \rho v} 
\label{app_cont}
\ee
\be
v\dvr = -{c_s^2\over \gamma}\left({1\over \rho}{d\rho\over dr}\right) -{1\over \gamma}{\dcsr}+f(r)-\dphr -{\dot m v \over \rho}
\label{app_mom}
\ee
where $c_s^2=\gamma p/\rho$. Elimination of ${1\over \rho}{d\rho\over dr}$ from above two equations yields,
\be
v\dvr - {c_s^2 \over \gamma v} \dvr = -\frac{1}{\gamma}\dcsr + {c_s^2\over \gamma}\left({2\over r}-{\dot m \over \rho v}\right) - {\dot m v \over \rho} + f(r) -\dphr
\label{app_contmom}
\ee
from energy equation (Equation 3) we have,
\be
v\dvr + {1\over \gamma-1}\dcsr = {q\over \rho v}- {\epsilon(r)\dot m \over \rho v} + f(r) -{\dphr}
\label{app_en}
\ee
where $\epsilon(r) = \frac{v^2}{2} + \frac{c_s^2}{\gamma-1}$. Elimination of $\dcsr$ from Equation \ref{app_contmom} and \ref{app_en} results in,
\be
\frac{v^2-c_s^2}{v}\dvr = {c_s^2}\left({2\over r}-{\dot m \over \rho v}\right)  - (\gamma-1)\left(\frac{q}{\rho v} - \frac{\dot{m}\epsilon(r)}{\rho v}\right)  +f(r)-\dphr - {\dot m v \gamma\over \rho}
\label{app_en_mom}
\ee
Next we introduce the Mach number $\mach=v/c_s$. We can write the following relations,

\begin{eqnarray}
v\dvr &=& \frac{c_s^2}{2}\dMr + \frac{\mach^2}{2}\dcsr \quad ;  \nonumber \\ 
{1\over v}\dvr &=& \frac{1}{2\mach^2}\dMr + \frac{1}{2c_s^2}\dcsr 
\end{eqnarray}
Using these relations in Equation \ref{app_en} and \ref{app_en_mom} we obtain,
\be
\dcsr = \frac{{q\over \rho v}- {\epsilon(r)\dot m \over \rho v} + f(r)-\dphr  - \frac{c_s^2}{2}\dMr}{{\mach^2\over 2} + {1\over \gamma-1}}
\label{app_feed_1}
\ee
and
\be
c_s^2\frac{\mach^2-1}{2\mach^2}\dMr + \frac{\mach^2-1}{2}\dcsr = {c_s^2}\left({2\over r}-{\dot m \over \rho v}\right)  - (\gamma-1)\left(\frac{q}{\rho v} - \frac{\dot{m}\epsilon(r)}{\rho v}\right)  +f(r)-\dphr - {\dot m v \gamma\over \rho}
\label{app_feed_2}
\ee
Substituting Equation \ref{app_feed_1} in \ref{app_feed_2} we get the following wind equation,
\be
\frac{\mach^2-1}{\mach^2(\mach^2(\gamma-1)+2)}\dMr = {2\over r}-(1+\gamma\mach^2){\dot m \over \rho v}-\frac{\dot m(1+\gamma\mach^2)}{2 \rho v}\left(\frac{\dot E/\dot M}{\epsilon(r)}-1\right) + \frac{(\gamma+1)\left[f(r)-\dphr\right]}{2(\gamma-1)\epsilon(r)}
\label{central_app}
\ee
 Where $\dot{M}=\dot{m}V$ and $\dot{E}=qV$ in which $V$ is the volume of injection region. This is the general form of wind equation with contributions from energy and mass injection, gravity and external driving force. 
\section{Appendix B: subsonic part of SN\lowercase{e} \& AGN driven wind}
Here we show the subsonic part of the solution for SNe and AGN driven wind. Apart from the energy and mass injection we take $f(r)=\Gamma G \mbh/r$ and $\Phi(r) = \Phi_\bullet(r) = -G\mbh/r$. 
From continuity equation we obtain $\rho v r^2 = \dot m r^3/3$. If we substitute this in energy equation we get,
\be
{d\over dr}\left[{\dot m r^3\over 3}\left(\frac{v^2}{2}+{c_s^2\over \gamma - 1}\right)\right]+ {\dot m r^3\over 3}\left({(1-\Gamma)G\mbh\over r^2}\right) + \left[{\dot m r^3\over 3}{d \Phi_{\rm NFW}\over dr}\right] =   {\dot{E}\ r^2\over V} 
\ee
In the subsonic regime $r<<r_s$. Therefore in this case the third term on left hand side which is due to NFW gravity can be neglected  as shown below,
\be
\left.{\dot m r^3\over 3}{d \Phi_{\rm NFW}\over dr}\right|_{r<<r_s} = \left.{-\dot m G M_h r\over 3\ [\ln(1+c)-c/(1+c)] }\left({r/r_s \over 1+r/r_s}-\ln(1+r/r_s)\right)\right|_{r<<r_s} \approx 0
\ee
Neglecting the third term, (${\dot m r^3\over 3}{d \Phi_{\rm NFW}\over dr}\approx 0$) and integrating from zero to  $r$ we get,
\begin{eqnarray}
&&\left.\left[{\dot m r^3\over 3}\left(\frac{v^2}{2}+{c_s^2\over \gamma - 1}\right)\right]\right|_0^r+\left.\left({\dot m(1-\Gamma)G\mbh r^2\over 6}\right)\right|_0^r =   \left.{\dot{E}\over V}{r^3\over 3}\right|_0^r \nonumber \\ 
\Rightarrow&& \epsilon(r) = {\dot E\over\dot M}- (1-\Gamma)\vbh^2{R\over r}= 2v_\star^2 - (1-\Gamma)\vbh^2{R\over r}
\label{solo}
\end{eqnarray}
where we have defined $\vbh = \sqrt{G\mbh/2R}$,  a velocity characteristic of the black hole. At $r=R$ we obtain,
\be
\epsilon(R) = {\dot E \over \dot M} - (1-\Gamma)\vbh^2 = 2v_\star^2  - (1-\Gamma)\vbh^2 = 2v_{\rm crit}^2
\label{agn:energy}
\ee 
where $v_{\rm crit}$ is the velocity at the critical point.
Substituting the expression for  $\epsilon(r)$ and also  $f(r) = \Gamma G\mbh/r^2$, $\Phi = \Phi_\bullet +\Phi_{\rm NFW}$, $\gamma=5/3$  in wind equation \ref{central} of the main text, we get,
\be
\frac{\mach^2-1}{\mach^2(\mach^2(2/3)+2)}\dMr = {2\over r}-{3+5\mach^2 \over r}-\frac{(3+5\mach^2)}{2\ r}\left(\frac{{(1-\Gamma)R\ \vbh^2\over2\ r\ v_\star^2}}{1-{(1-\Gamma)R\ \vbh^2\over2\ r\ v_\star^2}}\right) - 2\left(\frac{{(1-\Gamma)R\ \vbh^2\over r^2\ v_\star^2}}{1-{(1-\Gamma)R\ \vbh^2\over2\ r\ v_\star^2}}\right) 
\label{agnsub}
\ee
Solution of this equation gives the mach number and velocity in the subsonic part of the wind. We can clearly see from the above equation that for $\vbh=0$ which will mean the AGN is not present, the subsonic part of the CC85 solution is recovered. Also, it is recovered when $\Gamma=1$ because in that case, although the AGN is present but the outward radiation force  cancels the inward gravitational force everywhere.

\section{Appendix C: Mach Number versus distance diagrams}
\label{lastapp}
\begin{figure*}[h]
    \centering
    \includegraphics[scale=0.5]{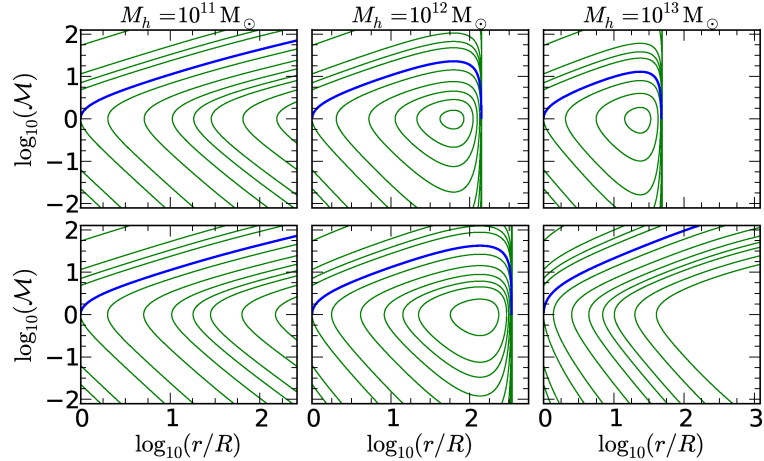}
    \caption{Mach number versus the radial distance for the winds from a Navarro-Frenk-White dark matter halo. Upper row of plots shows the effect of gravity of dark matter halo only and the lower row takes into account both the gravity due to halo and the effect of AGN. The three plots in each row are for three different halo masses. X-axis in each plot extends upto virial radius corresponding to the halo mass used. The thick blue line represents the solution with a critical point at $R=200$ pc,  which is used in this work. We have taken  $v_\star=180$ \kms.
    }
    \label{mach-contour}
\end{figure*}

Integral of the wind equation \ref{agn3} for the winds with NFW gravity and AGN is,
\be
\ln[\delta_>(\mach)] = 2 \ln [r] + 2 \ln \left[2v_{\rm crit}^2 -2(1-\Gamma)\vbh^2 + \Phi_{\rm NFW}(R) + 2(1-\Gamma) \vbh^2{R\over r}   -\Phi_{\rm NFW}(r) \right]  + const.
\ee
where $2v_{\rm crit}^2 = 2v_\star^2-(1-\Gamma)\vbh^2$. The above equation can equivalently be written as,
\be
\superdel \simeq A \ r^2 \left[2v_{\rm crit}^2 - 2(1-\Gamma){\vbh^2}\left(1-{R\over r} \right) -  2v_s^2 \left(1 - \frac{\ln(1+r/r_s)}{r/r_s}\right)\right]^2
\ee
where we have used the definition of function $\delta_>(\mach)$ given in \S 2.1. In this equation A is an arbitrary constant. 
If we set $A=1/(2Rv_{\rm crit})^2$, then for $r=R$ and $\mach=1$, both  LHS and RHS of the above equation become equal to 1. In Figure \ref{mach-contour} we plot the contours of Mach number versus the radius for three galaxies with a different halo masses. The upper three panels represent the winds without AGN ($\vbh=0$) and the lower three show the effect of the AGN momentum injection.  Different  contours in each panel correspond to a different value of the constant A. The thick blue contour in each panel represent the case with a critical point at $r=R=200$ pc and this one is used in the main text to calculate wind properties.

With the inclusion of NFW gravity one encounters a wall type of behaviour at a particular r. Beyond the wall the real and physical solutions are not possible. We would like to mention here that similar situation arises in adiabatic solar wind problem  as shown in Panel (c), Figure 2 of \citet{Holzer70}. The difference is that in the case of galaxy, the energy injection causes a critical point at R=200 pc.  
From Figure \ref{mach-contour} one can infer the interesting fact that the AGN is not able to drive the gas out of galaxy for intermediate halo masses but it can do so in high mass galaxies. 

\section{Appendix D: Possible effects of radiative cooling}
In this section we discuss the effect of radiation loss on the wind properties.
So far we have neglected the effect of radiative cooling, since it is known that the energy loss due to radiation  is too small to affect the dynamics of winds \citep{Grimes09}. However the cooling does effect the thermodynamics of the wind and can cause steeper fall in temperature as compared to the case with pure adiabatic expansion. We show this here with a simple case of supersonic escaping wind without gravity for which the wind equation with radiative cooling can be written as,
\begin{eqnarray}
\frac{\mach^2-1}{\mach^2(\mach^2(2/3)+2)}\dMr = &&{2\over r} + \frac{(\rho^2/m_p^2)\Lambda(T)}{2\rho v\ \epsilon(r)}(1+\gamma\mach^2)\nonumber \\
= && {2\over r} + \frac{3\dot M\Lambda(T)(3+5\mach^2 )(3+\mach^2)}{16\pi\ \epsilon(r)^2\mach^2}
\label{eq_cool}
\end{eqnarray}
where we have used $\rho=\dot M /(4\pi v r^2)$, $\gamma=5/3$ and $c_s^2 = 2\epsilon(r)/(\mach^2+3)$. Here $\Lambda(T)$ is the cooling rate in {\rm erg cm$^3$ s$^{-1}$}. We use $\Lambda\propto T^{0.8}$ for the temperature range $10^4<T<10^7$ and $\Lambda\propto T^2$ in the temperature range $10^4<T<10^5$ \citep{Suth93}.  To solve this equation we have to supply  $\epsilon(r)$ and $\dot M$. The specific enthalpy, $\epsilon(r)$ can be obtained from the integral  of energy equation which is given below, 
\begin{eqnarray}
&&\epsilon(r) = {v^2 \over 2} +{c_s^2 \over \gamma -1}= \frac{\dot E}{\dot M} - {1 \over \dot{M}}\int_R ^r {\rho ^2 \Lambda_{\rm cool} \over m_p} (4 \pi r^2) dr=\frac{\dot E}{\dot M} - {\dot E_{cool}(r)\over \dot M}
\end{eqnarray}
Considering the fact that total energy extracted by cooling over the entire path is $\sim10$\% of the adiabatic losses \citep{Wang95,Grimes09} we can neglect $\dot E_{cool}$ as compared to $\dot E$. Thus we get $\epsilon (r)=\dot E/\dot M = 2v_\star^2$. Using the $\epsilon(r)$ and a value for $\dot M$ in equation \ref{eq_cool}, we can solve it to obtain Mach number as a function of distance.

\begin{figure}[h]
   \centering
    \includegraphics[scale=0.5]{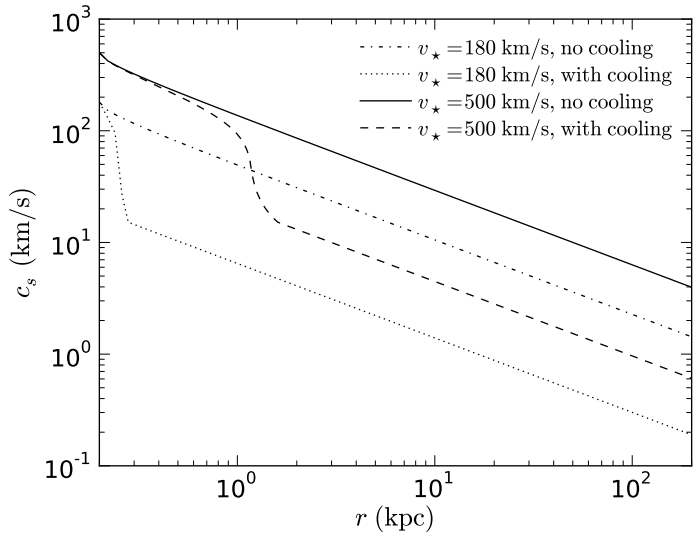}
    \caption{Sound speed for the wind with cooling is compared with the corresponding case without cooling. The Solid and dash dotted lines represent the case without cooling for quiescent mode and starburst mode respectively. Their counterparts with cooling are shown by  dashed and dotted lines.
    }
    \label{fig:cooling}
 \end{figure}

In Figure \ref{fig:cooling} we plot the sound speed  $c_s = [2\epsilon(r)/(\mach^2+3)]^{1/2}$ against r, where $\mach$ is obtained by solving equation \ref{eq_cool}. We have solved equation \ref{eq_cool} for the quiescent mode ($v_\star=180$ \kms) and the starburst mode ($v_\star=500$ \kms). We have used  $\dot M\sim 1\ \Msun$/yr and  $10\ \Msun$/yr as two fiducial value of mass loss rate for the quiescent mode and the starburst mode respectively. The sound speed corresponding to quiescent mode is shown by dotted line and that of the starburst mode by a dashed line.  We have also shown the corresponding cases without cooling using a dash-dotted and a solid line. 
By comparing the sound speeds with and without cooling, we find that  the cooling causes the temperature to decay more rapidly. Also cooling is effective at smaller distances ($r<10$ kpc) as the density and temperature are larger there. Wind speed is given by the relation $\epsilon(r) =  v^2/2 + c_s^2/(\gamma-1)\sim 2v_\star^2$. The terminal wind speed is obtained by neglecting the sound speed which is known to decrease with distance.  As the sound speed go down even more rapidly in case of cooling, therefore the wind speed beyond a distance of 10 kpc does not differ from the case without cooling and is given by, $v_{\rm wind} \approx 2v_\star$. 
 We would like to mention here that qualitatively similar and quantitatively more accurate result has also been obtained  by numerically solving the basic fluid equations with cooling for winds from individual star clusters in \cite{Silich04} and \cite{Tenorio07}. Whether the flow undergoing radiative energy losses can achieve a steady state, is also an interesting problem. We refer the reader to \citet{Silich03} where the  time dependent problem on a 2-D grid has been attempted and the result show that the flow can become steady after sufficient amount of time.

\section{Appendix E: Coupling between dust and gas}
The  equation for motion of dust grains acted upon by the radiation from AGN  can be written as,
 \be
 v_{\rm dr} {d v_{\rm dr}\over dr} = {\pi a^2 Q_{\rm rp} L \over 4 \pi r^2 c\ m_d}  - f_{\rm drag} -f_{\rm grav}
 \label{dust_mom}
 \ee
The first term on right hand side is the force of radiation per unit mass in which $Q_{\rm rp}$ is the radiation pressure mean efficiency, $a$ is the size  and $m_d$ is the mass of the dust grain. $f_{\rm grav}$ is the gravitational force per unit mass and $f_{\rm drag}$ is the drag force per unit mass due to the gas through which the dust grains drift with a velocity $v_{\rm dr}$. This drag force is given by,
\be
f_{\rm drag} = {\rho \pi a^2 v^2 \over m_d }
\ee
 where $\rho$ and $v$ is the density and velocity of the gas, and $a$ and $m_d$ is the size and mass of dust grain respectively. The drag force is a resistive force for the dust but the same drag is a driving force for the gas.
One has to simultaneously solve this equation along with the gas momentum equation to work out the general two phase structure of dusty winds. However many essential features can be captured in a so called 'single fluid approximation' where the dust grains attain a terminal drift speed and then the above written dust momentum equation need not be solved \citep{Simis03}. In that case one can substitute the entire radiation force into the gas momentum equation (Equation 2) because of the exact momentum coupling between dust and gas.  The dust grains receive momentum from photons and then pass it on to gas particles via collisions which further distribute it to other gas particles  \citep{Gilman72}.

The  momentum coupling and single fluid approximation can be applied if the dust grains attain a terminal drift speed quickly within a short distance once they start moving. We can now verify the validity of momentum coupling in the present case of a galactic outflow. Assuming a typical density profile $\rho = \rho_{o}(r_o/r)^2$ and neglecting the gravity in comparison to strong forces of radiation and drag, we can integrate equation \ref{dust_mom} to obtain the following solution (see also \citealt{Gilman72}),
\be
v_{\rm dr}^2(r) = v_{\rm T}^2\left[ 1- \exp\left(-{2 \ell \over r_o}(1-r_o/r) \right)\right]
\ee
where the terminal drift speed of dust grains is, $v_{\rm T}  = [Q_{\rm rp} L/4\pi c \rho_o r_o^2]^{1/2}$ with $l=\rho_o r_o^2 \pi a^2/m_d$ and $r_o$ is the launching radius. How quickly the  terminal drift speed is achieved, is decided by the value of the multiplicative factor $2l/r_o$ in the exponential. We can estimate $\ell$ for a typical grain size of $0.1\ \mu$m, and  a grain mass density of $3$ g cm$^{-3}$. The quantity $\rho_o r_o^2 = \rho_R R^2$, where $R=200$ pc is the critical point, $\rho_R = 1 $ m$_{\rm p}$ cm$^{-3}$ which is a typical value at the critical point in our wind models. Using these values we get $\ell \approx 5$ kpc. If the dust grains are launched at $r_o\sim 10$ pc, we have $2\ell/r_o\sim10^3$ and even for $r_o =100$ pc
we obtain $2\ell/r_o\sim10^2$. These large values of $2 \ell/r_o$ imply that the grains attain drift speed within a short distance. Once the grains are moving with the constant terminal drift speed ($v_{\rm T}$) the entire radiation force is transferred to the gas via the drag force. Therefore the exact coupling between the dust and the gas is justified. Thus the momentum injection force per unit mass of the gas is simply  given by $f(r)=n_d m_d f_{\rm drag}/\rho$. As the dust is moving with a terminal speed, it implies that the drag force is equal to the radiation force.  Therefore we can substitute the radiation force in place of $f_{\rm drag}$ to obtain the following expression for force on gas ($f(r)$),
\be
f(r) = {n_d Q_{\rm rp} \pi a^2 \over \rho} {L\over 4\pi r^2 c} = \kappa {L\over 4 \pi r^2 c}
\ee
where $\kappa$ is the opacity for a mixture of dust and gas.

\label{lastpage}


\end{document}